\documentclass[manuscript]{aastex}

\usepackage{emulateapj5}
\usepackage{graphicx}

\newcommand{{\galex}}{{\it GALEX}}
\newcommand{{\iue}}{{\it IUE}}
\newcommand{{\iras}}{{\it IRAS}}
\newcommand{{\akari}}{{\it AKARI}}

\slugcomment{}

\shorttitle{Reexamination of the IRX--$\beta$ Relation}
\shortauthors{Takeuchi et al.}

\begin{document}

\title{Reexamination of the Infrared Excess--Ultraviolet Slope Relation of \\ Local Galaxies}

\author{Tsutomu T.\ Takeuchi\altaffilmark{1},}
\altaffiltext{1}{Department of Particle and Astrophysical Science, Nagoya University, Furo-cho, Chikusa-ku,
  Nagoya 464--8602, Japan}
\email{takeuchi.tsutomu@g.mbox.nagoya-u.ac.jp}

\author{Fang-Ting, Yuan\altaffilmark{1}, Akira Ikeyama\altaffilmark{1}, Katsuhiro L.\ Murata\altaffilmark{1},}

\and

\vspace{-0.4cm}
\author{Akio K.\ Inoue\altaffilmark{2}}
\altaffiltext{2}{College of General Education, Osaka Sangyo University, 3-1-1 Nakagaito, Daito, Osaka 574-8530, Japan}

\begin{abstract}
The relation between the ratio of infrared (IR) and ultraviolet (UV) flux densities
(the infrared excess: IRX) and the slope of the UV spectrum ($\beta$) of 
galaxies plays a fundamental role in the evaluation of the dust attenuation of star forming 
galaxies especially at high redshifts. 
Many authors, however, pointed out that there is a significant dispersion
and/or deviation from the originally proposed IRX-$\beta$ relation 
depending on sample selection.
We reexamined the IRX-$\beta$ relation by measuring the far- and 
near-UV flux densities of the original sample galaxies with {\galex} and {\akari} imaging data,
and constructed a revised formula.
We found that the newly obtained IRX values were lower than the original relation because 
of the significant underestimation of the UV flux densities of the galaxies, caused by 
the small aperture of {\iue}, 
Further, since the original relation was based on \iras\ data which covered  
a wavelength range of $\lambda = 42\mbox{--}122\;\mu$m, 
using the data from \akari\ which has wider wavelength coverage toward
longer wavelengths, we obtained an appropriate IRX-$\beta$ relation with 
total dust emission (TIR): $\log \left( L_{\rm TIR}/L_{\rm FUV} \right) = 
\log \left[ 10^{0.4(3.06+1.58\beta )}-1 \right] +0.22$.
This new relation is consistent with most of the preceding results for samples selected at 
optical and UV, though there is a significant scatter around it.
We also found that even the quiescent class of IR galaxies follows this new relation, though
luminous and ultraluminous IR galaxies distribute completely differently as well known before. 
\end{abstract}


\section{Introduction}

The true star formation rate (SFR) of galaxies at various redshifts is one of the most fundamental
physical quantities to understand the formation and evolution of galaxies in the Universe.
Especially, the SFR density of galaxies in a cosmic volume is a very convenient
tool to explore the galaxy evolution.
The concept of the cosmic SFR density was originally introduced by \citet{tinsley80}, and subsequently
it has been measured observationally at various redshifts up to $z \sim 7\mbox{--}8$
\citep[e.g.,][among others]{lilly96,madau96,madau98,connolly97,bouwens04,bouwens07,
bouwens09}.

Since the SFR of galaxies is essentially determined by measuring observables related to
the ultraviolet (UV) flux from OB stars, like hydrogen recombination lines or forbidden lines
through ionizing UV photons, or UV continuum itself.
However, the largest uncertainty to determine the SFR from these observables is the dust
extinction. 
Since active star formation (SF) is always accompanied by dust production
because of various dust grain formation processes related to the final stage of 
stellar evolution \citep[e.g.,][]{dwek80,dwek98,nozawa03,nozawa10,takeuchi03,takeuchi05c,
draine03}, it is always necessary to deal with the energy ``hidden'' by dust.
The absorbed energy by dust is reradiated by dust grains at far-infrared (FIR) wavelengths
\citep[e.g.,][]{calzetti00,buat02}.
Now it is shown that the fraction of the hidden star formation is even dominant at $z \sim
1$ \citep{takeuchi05a}.
The most straightforward way to remedy the dust extinction problem would be to 
sum up the SFR calculated from observed (uncorrected) UV and FIR luminosities
\citep{buat07a,buat07b,takeuchi10}.
However, this is not always possible to obtain both restframe UV and FIR flux, especially
at high redshifts to date.

In general for Local galaxies, when dust grains in galaxies absorb UV light, 
shorter wavelength flux is more strongly attenuated than longer one, which causes
{\it reddening} of galaxy spectra\footnote{However note that if the extinction/attenuation curve is independent of
wavelength (gray extinction), reddening does not occur.}.
Then, the strength of reddening and absorption fraction of UV are expected to be related 
in some way. 
Evaluating this process, \citet{meurer99} (hereafter M99) proposed a formula to estimate 
dust attenuation from the slope of the ultraviolet (UV) spectrum, $\beta$. 
This is based on the tight relation between the infrared (IR) to ultraviolet flux ratio (traditionally 
referred to as the infrared excess: IRX) and $\beta$.
Since this method does not require any other wavelength data to obtain the value of attenuation, 
like IR emission, it was very much appreciated by researchers of high-redshift galaxies 
($z \sim 2\mbox{--}5$ or more).
This is because the dust emission of such high-$z$ galaxies is usually much more 
difficult to observe, and their UV spectra are observed at optical wavelengths 
thanks to the redshifts \citep[e.g.,][]{bouwens07,bouwens09}.
Thus, the IRX-$\beta$ relation is very popularly used in these studies up to now.

However, recently, subsequent studies have shown that the original formula of 
M99\nocite{meurer99} does not always fit to Local galaxies 
\citep[e.g.][]{buat05,seibert05,burgarella05a,buat07a,buat07b,cortese06,boissier07,bouquien09,takeuchi10,overzier11}.
Especially, rather quiescently star-forming galaxies tend to distribute significantly 
below the IRX-$\beta$ relation of M99.
In addition, another extreme categories of starburst galaxies, luminous and ultraluminous IR galaxies
(LIRGs and ULIRGs, respectively), occupy a region well above the M99 relation
\citep{goldader02,buat05,burgarella05b,takeuchi10}.
Also, it is reported that high-$z$ star-forming galaxies also tend to deviate from the M99
relation \citep[e.g.,][]{siana09,reddy10}.
Since the M99's IRX-$\beta$ relation was proposed for UV-luminous starburst galaxies, 
one may think it is not surprising that the M99 sample galaxies have different property from IR-selected
galaxies.
Indeed, some attempts have been done to explain {\it physically} the discrepancy and dispersion 
of the IRX-$\beta$ relation for various categories of galaxies 
\citep[e.g.][]{kong04,bouquien09,conroy10}.
However, since this relation is often applied to high-$z$ galaxies whose nature is quite unknown,
it is reasonable to examine the M99 relation according to their original
recipe for its firm application in future studies.

The original IRX-$\beta$ relation of M99 was obtained based on {\iras} and {\iue} data.
Though the latter was an epoch-making facility for the UV astronomical studies, it had
one observational limitation: 
the aperture of {\iue} is $10'' \times 20''$ ellipse with effective circle-equivalent { diameter of $14''$.}
As M99 discussed in their original work, the sample of starburst galaxies in M99\nocite{meurer99}
sometimes have large angular diameters (see Section~\ref{subsec:m99}). 
In contrast, the beam size of {\iras} is quite large, and most of the part of galaxies is covered 
by the {\iras} instrument.
This difference of apertures of {\iue} and {\iras} would cause a problem in the 
estimation of the IRX-$\beta$ relation since it is based on the ratio of the two.

Today, we have a much larger dataset open to public at UV bands provided by
{\galex}.
{\galex} (Galaxy Evolution Explorer) is a UV astronomical satellite launched in 2003 by NASA
\citep{morrissey07}\footnote{URL: http://www.galex.caltech.edu/.}. 
{\galex} has far-UV (FUV) and near-UV (NUV) bands, and is performing 
all-sky survey as well as deep surveys in some sky areas.
In this work, we reexamine the M99 relation by using the same galaxy sample as 
M99\nocite{meurer99}, with paying a particular attention to the aperture problem.
For this purpose, we use {\galex} NUV and FUV images for the M99's original sample
galaxies. 
Another important aspect of using \galex\ filters for estimating $\beta$ of the M99's
sample is that since the main target of the dust correction by the IRX-$\beta$ relation
is high-$z$ galaxies, for which usually only broadband photometric data are available.
Since \iue\ data are spectroscopic, it is crucial to recalibrate photometrically estimated
UV slope with directly measured one from \iue\ spectroscopic data (see Section~\ref{subsubsec:filter}).

It is also interesting and important to examine the IRX-$\beta$ relation from
the FIR side.
For this, we also have a large all-sky IR data obtained by {\akari}\footnote{The {\akari} 
all-sky survey point source catalogs at MIR and FIR can be retrieved from the URL: 
http://www.ir.isas.ac.jp/ASTRO-F/Observation/PSC/Public/.}.
{\akari} is an infrared astronomical satellite launched in 2006 by JAXA (Japan Aerospace
Exploration Agency).
{\akari} is equipped with two imaging instruments, a near- and mid-IR camera IRC \citep{onaka07}
and FIR surveyor FIS \citep{kawada07}, and a Fourier spectrograph \citep{kawada08}.
In this study, we make use of the data obtained by FIS.
The FIS has four photometric bands, {\it N60} ($50\mbox{--}80~\mu$m), {\it WIDE-S} ($60\mbox{--}110~\mu$m), 
{\it WIDE-L} ($110\mbox{--}180~\mu$m), and {\it N160} ($140\mbox{--}180~\mu$m).
Since {\akari} has longer wavelength bands than that of {\iras}, we can estimate the
total FIR emission from dust directly without {any} complicated conversion formula
\citep[e.g.,][]{dale01, dale02}.
Thus, with \galex\ and \akari\ data, we can obtain an unbiased IRX-$\beta$ relation for
the M99 sample. 

This paper is organized as follows: in Section~\ref{sec:sample}, we describe the data of
M99\nocite{meurer99} and measurements newly done in this study.
Section~\ref{sec:results}, we first show the result of the {\galex}-{\iras} IRX-$\beta$ relation. 
Then we examine this new relation, and discuss the origin of the discrepancy
between the M99 relation and recent observations from UV side.
Then, we examine the IR side with {\akari} FIS Bright Source Catalog (FISBSC). 
After considering all the measurement-related effects with \akari\ diffuse map, 
we propose a new IRX-$\beta$ relation for the UV-selected starbursts with {\galex} and {\akari}.
In Section~\ref{sec:discussion}, we further examine some properties related to the IRX-$\beta$ 
relation we obtained in this work.
First we estimate the bolometric correction for FIR by \iras\ and \akari\ flux
densities.
Then, we examined the behavior of the IRX as a function of aperture radius but both on the \galex\ and
\akari\ images. 
By this analysis, we can obtain the IRX only from the central starburst regions of
the original M99 sample.
After checking these properties, we finally compare our new relation to various previous results.
Section~\ref{sec:conclusion} is devoted to our conclusions.
We review the formulation of M99\nocite{meurer99} in Appendix~\ref{sec:meurer}.
In Appendix~\ref{sec:akari_diffuse}, we evaluate the aperture 
effect in \akari\ FISBSC flux densities with \akari\ FIS diffuse map.

\section{Sample and Data Analysis}\label{sec:sample}

\subsection{Original data of M99}\label{subsec:m99}

First, we describe the original data of M99\nocite{meurer99}.
\citet{kinney93} has constructed a UV SED atlas of galaxy sample of 143 spiral, irregular, blue compact 
dwarfs, Seyfert 2, and starburst galaxies with {\iue} spectrum archive.

M99\nocite{meurer99} selected a subsample of very actively star-forming 
galaxies from this database which consisted of starburst nucleus, blue compact, 
and blue compact dwarfs.
Further, in order to reduce the effect of small aperture of {\iue}, 
M99\nocite{meurer99} have selected galaxies whose surface brightness 
distribution is strongly concentrated in the central small region. 
{}To construct this sample, they used a criterion for an optical diameter 
$D_{25}<4'$ ($D_{25}$ is measured at the $25\; \mbox{mag\,arcsec}^{-2}$ isophote).
We will revisit this issue in the following.

This further selection resulted in 57 starbursts in the {\iue} sample. 
M99 used 44 galaxies which were observed at {\iras} $60~\mu$m and $100~\mu$m
bands to make the IRX$-\beta$ plot and obtained the Meurer relation.
The sample and its property are summarized in Table~\ref{table:meurer}.

\begin{figure*}[t]
\centering\includegraphics[width=0.7\textwidth]{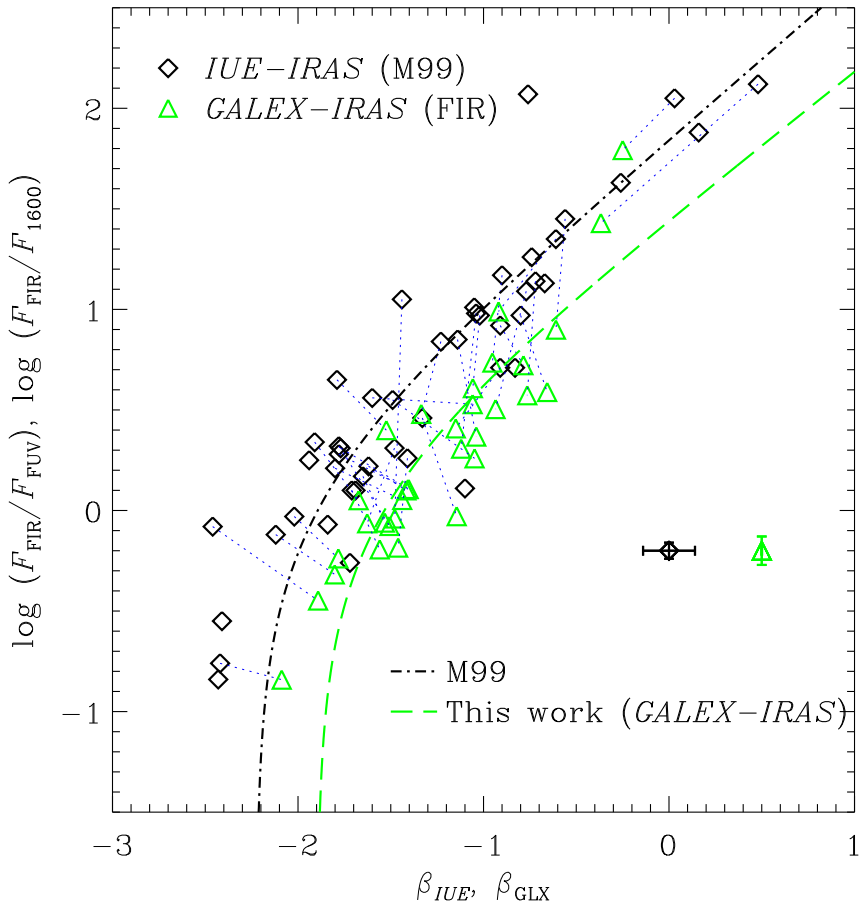}
\caption{
The IRX-$\beta$ relation obtained with {\galex} and {{\iras}}.
Open squares represent the original data of (\citealt{meurer99}: M99), 
and filled circles are the new measurements from {\galex} images in this work.
A pair of symbols connected by dotted lines represents the same galaxy.
Typical uncertainty for each measurement is indicated on the lower-left.
For the original {\iue} measurements of M99, the abscissa is the 
UV slope of the {\iue} spectra of the sample $\beta$, and for the {\galex} measurements,  
it is the slope defined by the UV color defined by the FUV and NUV photometry 
by {\galex} [eq.~(\ref{eq:beta_galex})], $\beta_{\rm GLX}$ \citep{kong04}.
As for the ordinate, it is the ratio between the FIR flux $F_{\rm FIR}$ [eq.~(\ref{eq:fir_helou})]
and the 1600~\AA\ flux $F_{1600}$ [eq.~(\ref{eq:f1600})], $F_{\rm FIR}/F_{1600}$
(referred to as the IRX) for the original M99 data, and the flux ratio between
$F_{\rm FIR}$ and the {\galex} FUV-band flux $F_{\rm FUV}$, $F_{\rm FIR}/F_{\rm FUV}$.
}\label{fig:irx_beta_galex_iras}
\end{figure*}

\subsection{New measurement of UV flux by GALEX images}\label{subsec:galex_phot}

In order to examine the possible aperture effect of {\iue}, we remeasured the
UV flux densities from {\galex} images.
As already mentioned, \galex\ has two UV bands, FUV ($\lambda = 1344\mbox{--}1786$~\AA) and 
NUV ($\lambda = 1771\mbox{--}2831$~\AA).
The \galex\ images are available through MAST\footnote{MAST URL: http://galex.stsci.edu/GR6/.}. 

We have made a FUV and NUV photometry of the sample galaxies as follows:
\begin{enumerate}
\item Cut out a $30' \times 30'$ square subimage from {\galex} AIS images
  around each galaxy 
\item Select a subimage with the largest exposure time when multiple
  observations were available.
\item Measure FUV and NUV flux densities.  
The NUV image is taken as the reference.
\end{enumerate}
In this study, it is important to mention that almost all of the sources are resolved by {\galex}.
Then, they very often appear as an assembly of small bright patchy regions, 
and the {\galex} pipeline misidentifies these fragments as individual
objects. 
This is referred to as shredding.
We must deal with the shredding to obtain sensible flux density 
measurements for nearby extended galaxies.  
For this purpose, we have used an IDL software package developed 
by ourselves.  
This software performs aperture photometry in the NUV sub-image 
using a set of elliptical apertures.
Total flux density is calculated within the aperture corresponding to
the convergence of the growth curve.  
The sky background is measured by combining several 
individual regions around the source.  
NUV and FUV flux densities are corrected for Galactic extinction using 
the Schlegel map \citep{schlegel98} and the Galactic
extinction curve of \citet{cardelli89}.  
A detailed description of the photometry process can be found in 
\citet{iglesias06}.
 
This was already used for previous studies
\citep{iglesias06,buat07a,takeuchi10,sakurai12}, and its performance is carefully checked and
established.  
The new measurements are tabulated as $\log L_{\rm NUV}$ and $\log L_{\rm NUV}$
in Table~\ref{table:meurer}.

\begin{table*}[tp]
\begin{center}
\caption{Sample of \citet{meurer99} with new measurements.}\label{table:meurer}
{
\tiny
\scalebox{0.71}[0.71]
{
\bigskip
\begin{tabular}{lccccccccccc}\hline\hline \\
\bigskip
Name&$\beta$\tablenotemark{a}&$\beta_{\rm GLX}$\tablenotemark{b} & $\log L_{1600}$\tablenotemark{c}&
$\log L_{\rm FUV}$\tablenotemark{d}&
$\log L_{\rm NUV}$\tablenotemark{e}& $\log L_{\rm FIR}$\tablenotemark{f}&$\log L_{\rm TIR}$\tablenotemark{g} &
Major axis\tablenotemark{h} &Minor axis\tablenotemark{i} &$z$\tablenotemark{j}&Type\tablenotemark{k} \\
&	&	&[$L_{\odot}$]&[$L_{\odot}$]&[$L_{\odot}$]&[$L_{\odot}$]&[$L_{\odot}$]&[arcsec]&[arcsec]&&\\
\hline
&&&&&&&&&&&\\
NGC4861 & $-$2.46 & $-$1.89 & 8.76 & 9.13 & 8.97 & 8.68 & 8.50 & 180.00 & 46.15 & 0.0028 & SB \\
IZw18 & $-$2.43 & $-$2.09 & ... & 8.05 & 7.86 & ... & ... & 31.50 & 14.70 & 0.0025 & BCG \\
NGC1705 & $-$2.42 & $-$2.09 & 8.88 & 8.95 & 8.75 & 8.12 & 8.50 & 76.50 & 73.44 & 0.0021 & SA0 \\
Mrk153 & $-$2.41 & $-$1.78 & 9.20 & 9.38 & 9.24 & 8.65 & ... & 27.00 & 18.69 & 0.0080 & Scp \\
Tol1924$-$416 & $-$2.12 & $-$1.80 & 9.66 & 9.86 & 9.72 & 9.54 & 9.53 & 30.00 & 20.00 & 0.0095 & pec \\
UGC9560 & $-$2.02 & $-$1.78 & 8.57 & 8.77 & 8.63 & 8.54 & 8.77 & 42.00 & 27.00 & 0.0039 & pec \\
Mrk66 & $-$1.94 & $-$1.41 & 9.60 & 9.86 & 9.79 & 9.85 & ... & 22.50 & 16.36 & 0.0210 & BCG \\
NGC3991 & $-$1.91 & $-$1.67 & 9.68 & 9.96 & 9.84 & 10.02 & 10.11 & 61.50 & 28.38 & 0.0106 & PEC \\
NGC3738 & $-$1.89 & $-$1.49 & ... & 7.94 & 7.85 & ... & 7.86 & 76.50 & 62.16 & 0.0008 & Irr \\
UGCA410 & ... & ... & ... & ... & ... & ... & ... & ... & ... & 0.0022 & BCDG \\
Mrk357 & $-$1.80 & $-$1.44 & 10.66 & 10.81 & 10.73 & 10.87 & 11.03 & 25.50 & 19.83 & 0.0529 & Pair \\
NGC3353 & $-$1.79 & $-$1.52 & 8.50 & 8.73 & 8.63 & 9.15 & 9.23 & 61.50 & 45.32 & 0.0031 & Irr \\
MRK54 & $-$1.78 & $-$1.41 & 10.54 & 10.71 & 10.64 & 10.82 & 10.77 & 43.50 & 23.20 & 0.0449 & Sc \\
NGC1140 & $-$1.78 & $-$1.51 & 9.07 & 9.46 & 9.37 & 9.39 & 9.72 & 81.00 & 47.03 & 0.0050 & pec \\
Mrk36 & $-$1.72 & $-$1.32 & 7.75 & 8.47 & 8.41 & 7.49 & 8.40 & 150.00 & 28.68 & 0.0022 & BCG \\
MCG6$-$28$-$44 & $-$1.77 & $-$1.40 & 10.51 & 10.71 & 10.64 & 10.82 & 10.77 & 43.50 & 23.20 & 0.0449 & Sc \\
NGC1510 & $-$1.71 & $-$1.56 & 8.27 & 8.54 & 8.44 & 8.37 & 8.49 & 63.00 & 51.88 & 0.0030 & SA0 \\
MRK19 & ... & ... & ... & ... & ... & ... & 9.74 & ... & ... & 0.0138 & Sa \\
NGC4214 & $-$1.69 & $-$1.56 & ... & 9.01 & 8.91 & ... & 9.05 & 337.50 & 225.00 & 0.0010 & IAB \\
NGC4670 & $-$1.65 & $-$1.53 & 8.84 & 9.08 & 8.98 & 9.01 & 9.36 & 69.00 & 53.32 & 0.0036 & SB0 \\
NGC1800 & $-$1.65 & $-$1.63 & ... & 8.62 & 8.51 & ... & 8.31 & 67.50 & 35.10 & 0.0027 & IB \\
UGC5720 & $-$1.62 & $-$1.42 & 8.93 & 9.05 & 8.97 & 9.15 & 9.66 & 34.50 & 31.85 & 0.0048 & Im \\
UGC5408 & $-$1.60 & $-$1.06 & 8.98 & 9.01 & 9.00 & 9.54 & 9.78 & 34.50 & 26.83 & 0.0100 & E/S0 \\
NGC7673 & $-$1.50 & $-$1.27 & ... & 9.98 & 9.93 & ... & 10.41 & 51.00 & 45.00 & 0.0114 & SAc \\
NGC3125 & $-$1.49 & $-$1.05 & 8.71 & 8.99 & 8.99 & 9.26 & 9.52 & 40.50 & 29.25 & 0.0037 & BCdG \\
Haro15 & $-$1.48 & $-$1.63 & 9.96 & 10.33 & 10.22 & 10.27 & 10.44 & 52.50 & 43.24 & 0.0214 & SB0 \\
NGC2537 & $-$1.44 & $-$1.48 & 7.29 & 8.36 & 8.27 & 8.34 & 8.53 & 66.00 & 58.38 & 0.0014 & SB(s)m \\
UGC3838 & $-$1.41 & $-$1.46 & 8.97 & 9.41 & 9.33 & 9.23 & 8.30 & 42.00 & 24.71 & 0.0102 & Sd \\
NGC7793 & $-$1.34 & $-$1.58 & ... & 8.94 & 8.84 & ... & 9.08 & 354.00 & 200.26 & 0.0008 & SA(s)d \\
NGC5253 & $-$1.33 & $-$1.18 & ... & 8.95 & 8.92 & ... & 9.26 & 123.00 & 69.62 & 0.0014 & Im \\
NGC7250 & $-$1.33 & $-$1.14 & 8.71 & 9.20 & 9.17 & 9.17 & 9.34 & 49.50 & 19.04 & 0.0039 & Sdm \\
Mrk542 & $-$1.32 & $-$0.99 & ... & 9.64 & 9.64 & ... & 10.27 & 27.00 & 23.14 & 0.0245 & Im \\
NGC7714 & $-$1.23 & $-$1.34 & 9.54 & 9.90 & 9.84 & 10.38 & 10.48 & 40.50 & 24.92 & 0.0093 & SB(s) \\
Mrk487 & ... & ... & ... & ... & ... & ... & ... & ... & ... & 0.0022 & BCDG \\
NGC3049 & $-$1.14 & $-$1.04 & 8.44 & 8.93 & 8.92 & 9.29 & 9.38 & 90.00 & 45.00 & 0.0049 & SB(rs) \\
UGC6456 & $-$1.10 & $-$1.87 & 5.98 & 6.64 & 6.49 & 6.09 & ... & 37.50 & 26.47 & 0.0003 & pec \\
NGC3310 & ... & ... & ... & ... & ... & ... & 9.41 & ... & ... & 0.0033 & SAB(r) \\
NGC5996 & $-$1.04 & $-$1.12 & 9.23 & 9.90 & 9.88 & 10.21 & 10.38 & 76.50 & 42.08 & 0.0110 & SBc \\
NGC4385 & $-$1.02 & $-$1.06 & 8.85 & 9.20 & 9.19 & 9.82 & 10.01 & 73.50 & 36.75 & 0.0071 & SB(rs) \\
Mrk499 & $-$1.02 & $-$0.82 & ... & 9.88 & 9.92 & ... & 10.35 & 31.50 & 27.56 & 0.0260 & Im \\
NGC5860 & $-$0.91 & $-$0.78 & 9.32 & 9.51 & 9.55 & 10.24 & 10.38 & 43.50 & 35.59 & 0.0180 & E/S0 \\
IC1586 & $-$0.91 & $-$1.15 & 9.36 & 9.65 & 9.63 & 10.07 & 10.20 & 37.50 & 33.33 & 0.0194 & BCG \\
NGC2782 & $-$0.90 & $-$0.95 & 9.12 & 9.55 & 9.56 & 10.29 & 10.49 & 82.50 & 61.21 & 0.0085 & SAB(rs) \\
ESO383$-$44 & ... & ... & ... & ... & ... & ... & 10.12 & ... & ... & 0.0126 & SA(s) \\
NGC5236 & $-$0.83 & $-$0.80 & ... & 9.98 & 10.01 & ... & 10.68 & 352.50 & 320.11 & 0.0017 & SAB(s) \\
NGC2415 & $-$0.80 & $-$0.66 & 9.64 & 10.01 & 10.08 & 10.61 & 10.44 & 34.50 & 34.50 & 0.0126 & Im \\
MCG$-$01$-$30$-$33 & $-$0.77 & $-$0.93 & ... & ... & ... & ... & ... & 57.00 & 21.92 & ...00 & SB(s)b \\
NGC1614 & ... & ... & ... & ... & ... & ... & 11.42 & ... & ... & 0.0159 & SB(s)c \\
NGC6217 & ... & ... & ... & ... & ... & ... & 10.03 & ... & ... & 0.0045 & SB(rs) \\
NGC6052 & $-$0.72 & $-$0.76 & 9.55 & 10.11 & 10.15 & 10.69 & 10.89 & 55.50 & 45.71 & 0.0158 & Sc \\
NGC4500 & ... & ... & ... & ... & ... & ... & 10.37 & ... & ... & 0.0104 & SB(s)a \\
IC214 & $-$0.61 & $-$0.92 & 9.83 & 10.18 & 10.20 & 11.18 & 11.30 & 43.50 & 31.90 & 0.0302 & Irr \\
NGC3504 & $-$0.56 & $-$0.61 & 8.77 & 9.31 & 9.38 & 10.22 & 10.44 & 102.00 & 63.41 & 0.0051 & SAB(s) \\
NGC4194 & ... & ... & ... & ... & ... & ... & 10.67 & ... & ... & 0.0083 & BCG \\
NGC2798 & 0.03 & $-$0.25 & 8.22 & 8.48 & 8.62 & 10.27 & 10.26 & 28.50 & 23.75 & 0.0058 & SB(s) \\
NGC3256 & ... & ... & ... & ... & ... & ... & 11.29 & ... & ... & 0.0094 & Pec \\
NGC7552 & 0.48 & $-$0.37 & 8.66 & 9.35 & 9.47 & 10.78 & 10.78 & 111.00 & 62.83 & 0.0054 & SB(s) \\
\hline
\end{tabular}
}
}
\end{center}
\end{table*}

\setcounter{table}{0}
\begin{table*}[t]
\tablenotetext{\rm a}{: $f_{\lambda}\propto\lambda^{\beta}$\citep{meurer99}.}
\tablenotetext{\rm b}{: $\beta_{\rm GLX}$ defined by \citet{kong04}.}
\tablenotetext{\rm c}{: {\iue} 1600\AA\ luminosity ~\citep{meurer99}.}
\tablenotetext{\rm d}{: {\galex} FUV luminosity.}
\tablenotetext{\rm e}{: {\galex} NUV luminosity.}
\tablenotetext{\rm f}{: {\iras} FIR luminosity ~\citep{meurer99}.}
\tablenotetext{\rm g}{: {\akari} total luminosity.}
\tablenotetext{\rm h}{: Major axis of isophotal ellipse.}
\tablenotetext{\rm i}{: Minor axis of isophotal ellipse.}\tablenotetext{\rm j}{: Redshift taken from NED.}
\tablenotetext{\rm k}{: Morphological type taken from NED.}
\end{table*}

\begin{figure*}[t]
\centering\includegraphics[width=0.7\textwidth]{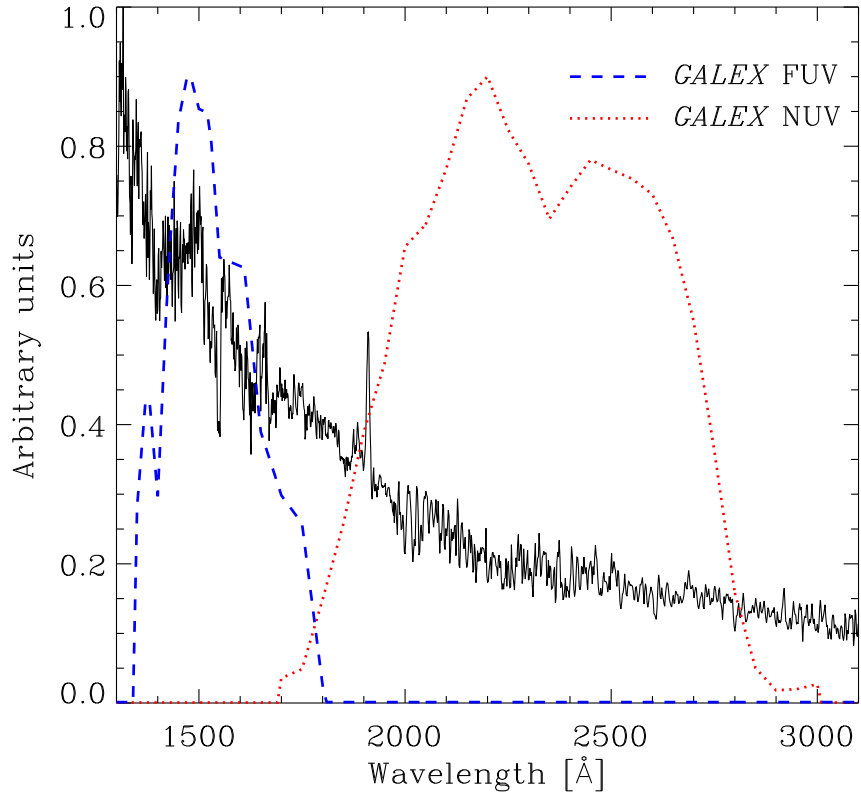}
\caption{Example of {\iue} galaxy spectrum in \citet{meurer99} sample (NGC4861)
with \galex\ FUV and NUV filter response functions. 
Solid curve represents the observed spectrum of NGC4861 by {\iue}.
Dashed and dotted lines are response functions of \galex\ FUV and NUV filters,
respectively.
All are arbitrarily normalized. 
}\label{fig:NGC4861}
\end{figure*}
\begin{figure*}[t]
\centering\includegraphics[width=0.4\textwidth]{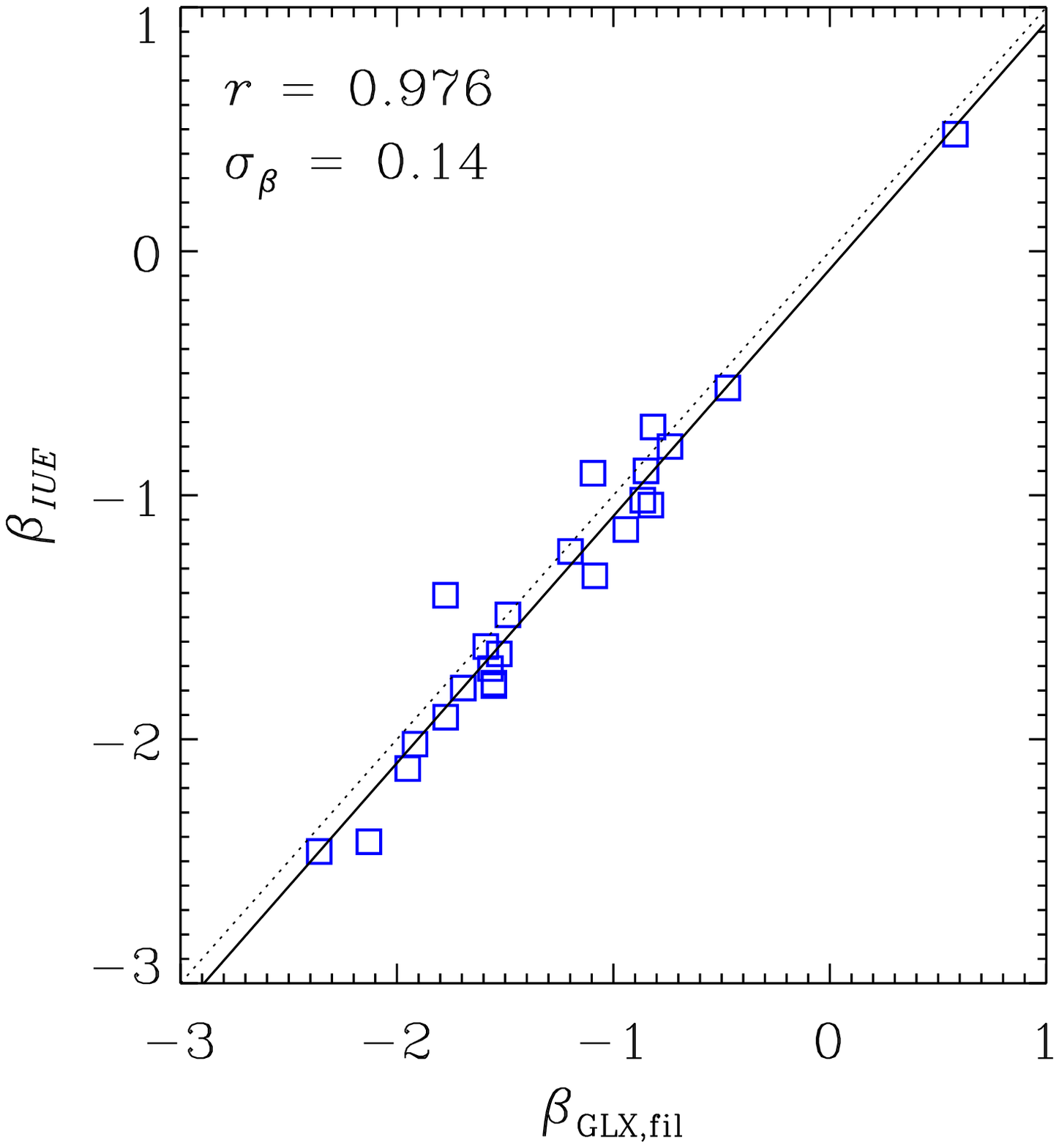}
\centering\includegraphics[width=0.4\textwidth]{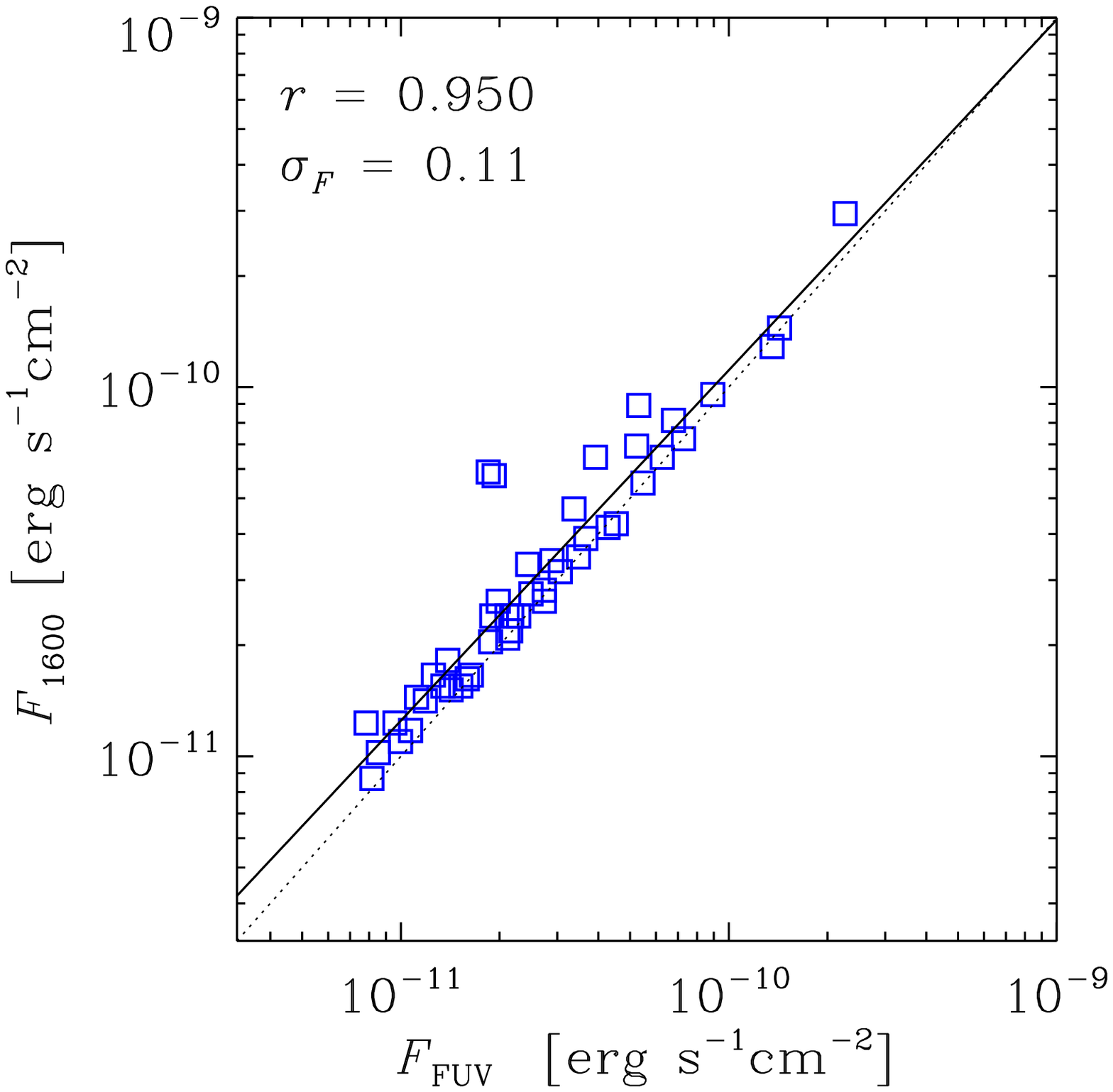}
\caption{
Comparison of the UV spectral slope $\beta$ and the 1600~\AA\ flux density before and
after transmitting the \galex\ filters. 
Left: the $\beta$ derived by {\galex} FUV and NUV photometry and that 
directly measured by \iue\ spectra; 
Right: the 1600~\AA\ flux density through the {\galex} FUV filter and that directly measured
by {\iue}. 
Solid lines represent the linear fit to the data, and diagonal dotted lines in both panels show 
$y = x$.
}\label{fig:iue_galex_filter}
\end{figure*}

\section{Results}\label{sec:results}

\begin{figure*}[tb]
\centering\includegraphics[width=0.7\textwidth]{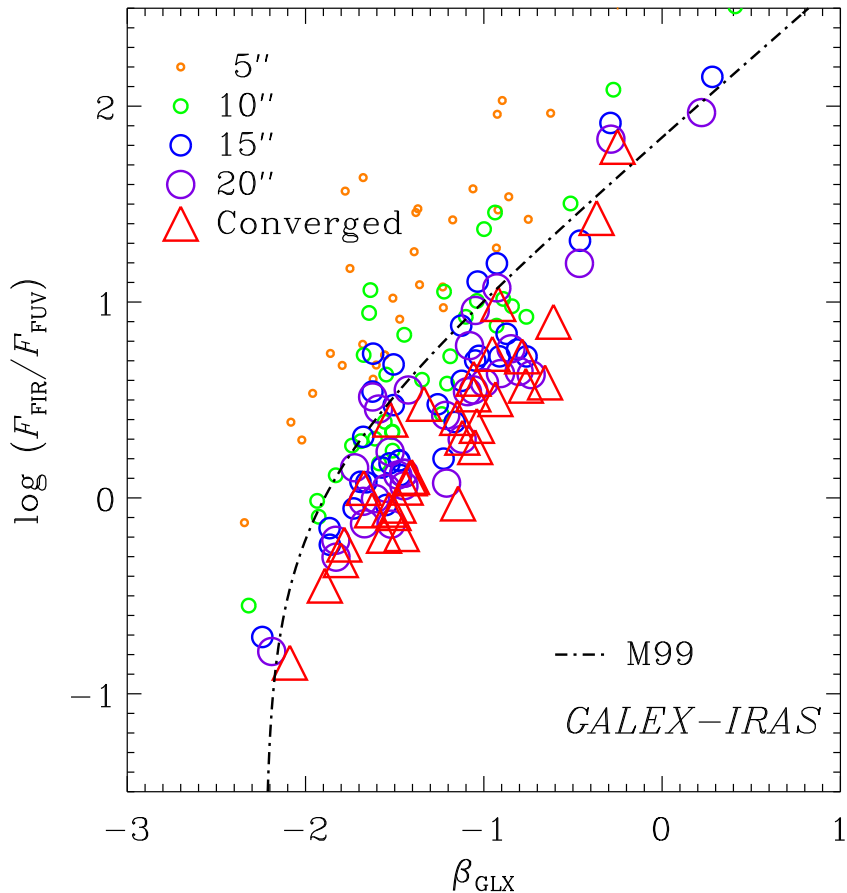}
\caption{
Behavior of the IRX-$\beta$ relation with the aperture radius on \galex images.
The semi-major axis radius of the photometry ellipse is set to be $5''$, $10''$, $15''$, $20''$ from 
smaller to larger open circles.
We also put the converged values of the growth curve of flux given by our software
by open triangles.
}\label{fig:IRX_beta_all}
\end{figure*}

\begin{figure*}[p]
\centering\includegraphics[width=0.35\textwidth]{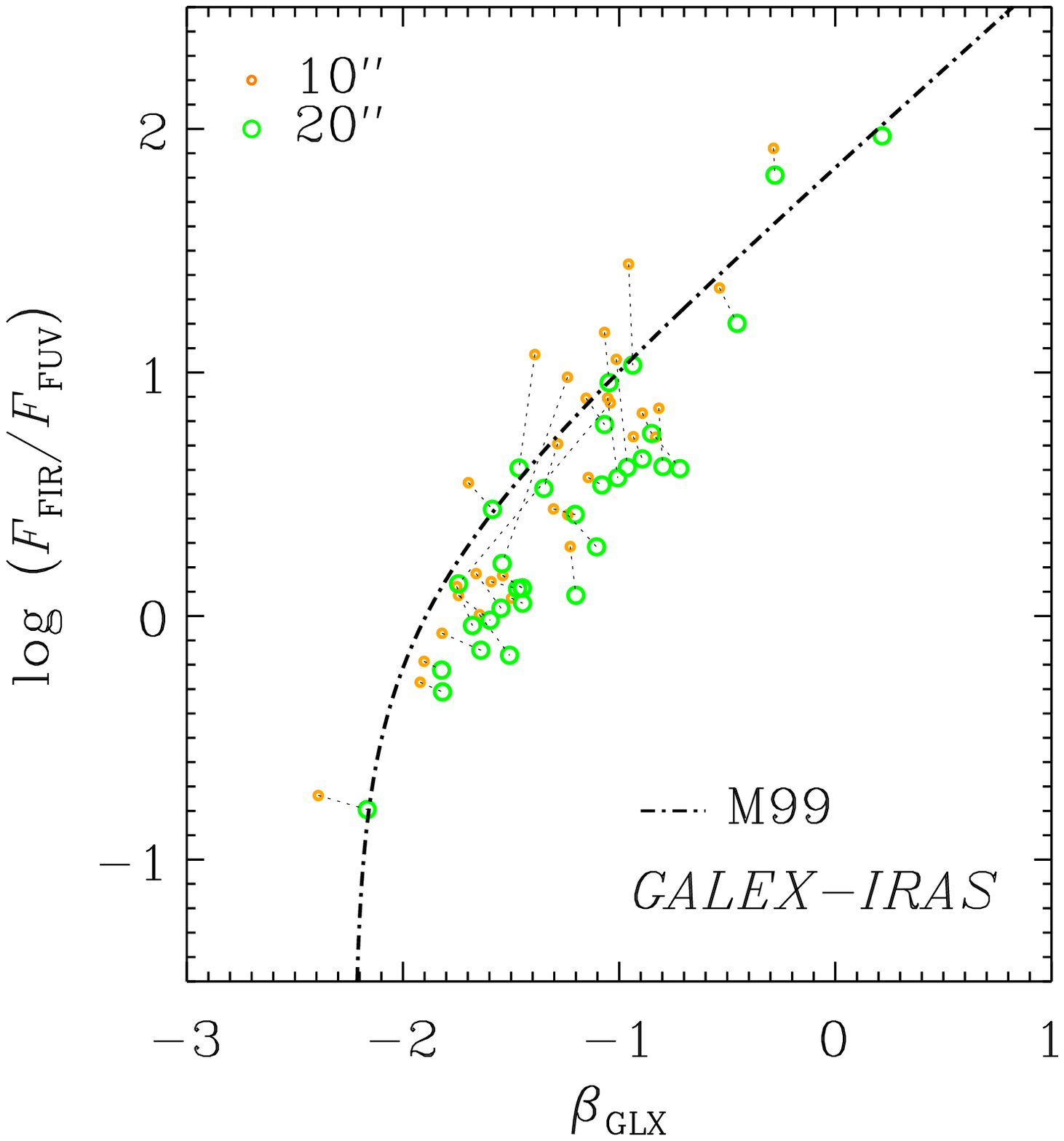}
\centering\includegraphics[width=0.35\textwidth]{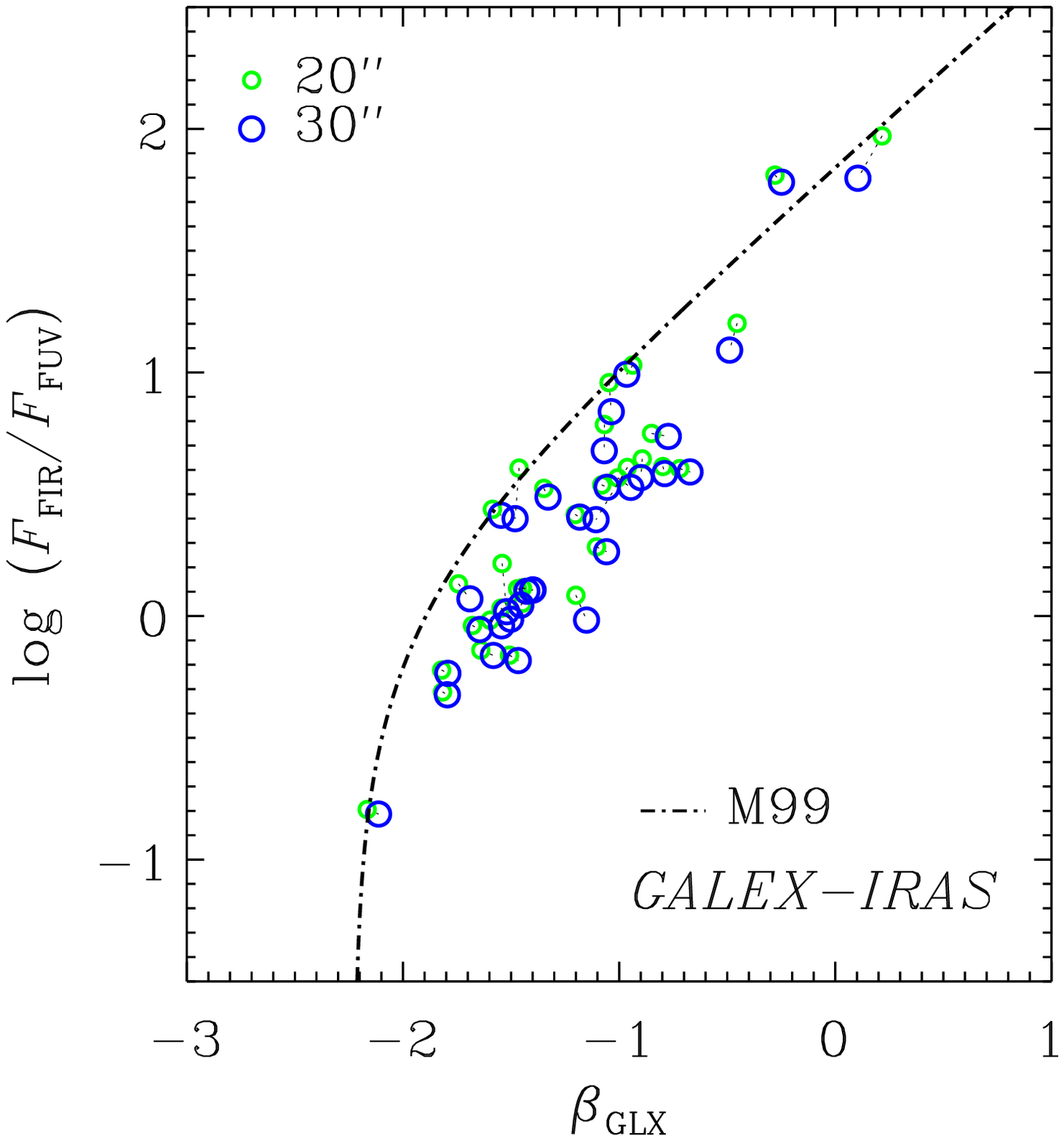}
\centering\includegraphics[width=0.35\textwidth]{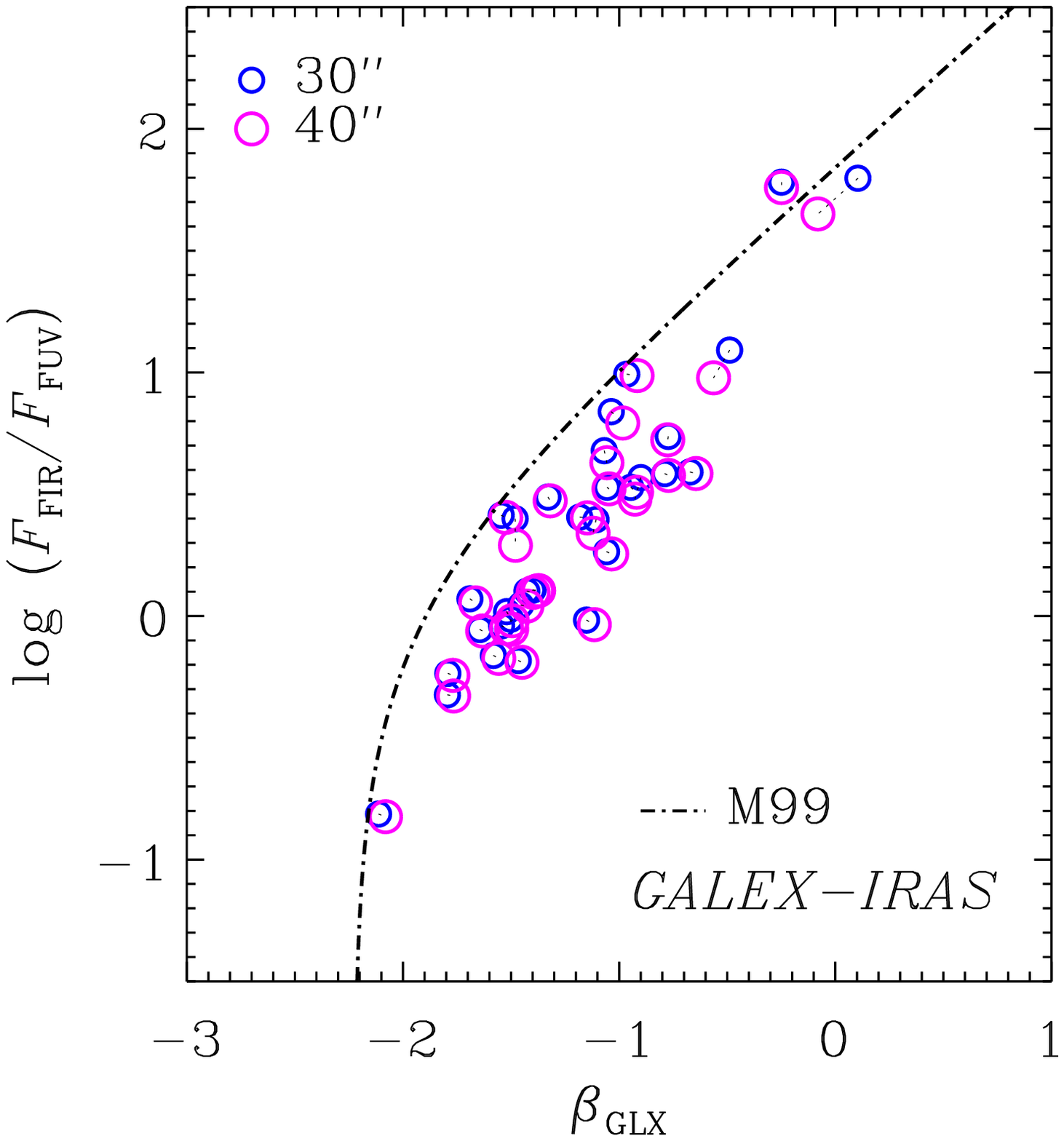}
\centering\includegraphics[width=0.35\textwidth]{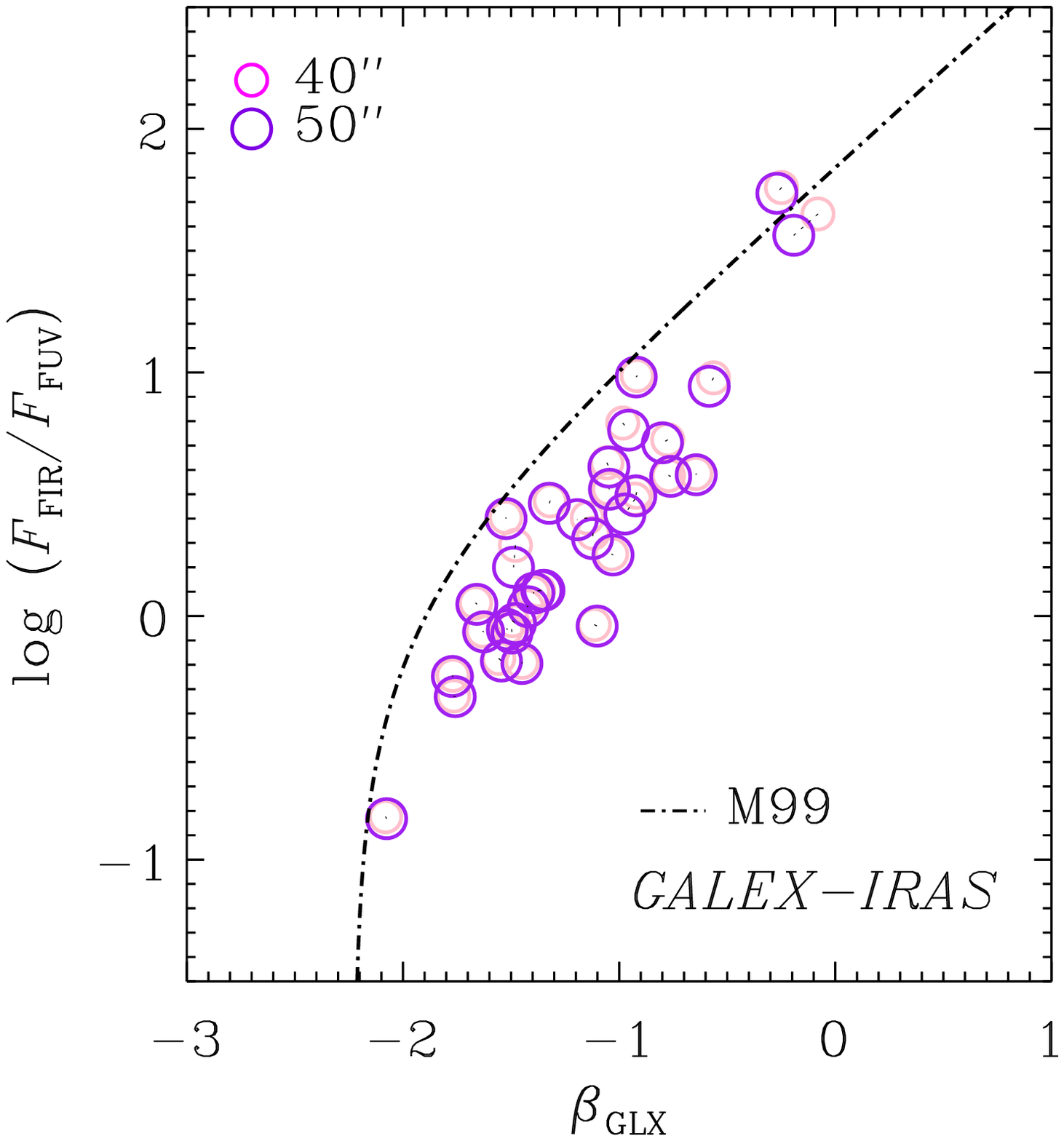}
\centering\includegraphics[width=0.35\textwidth]{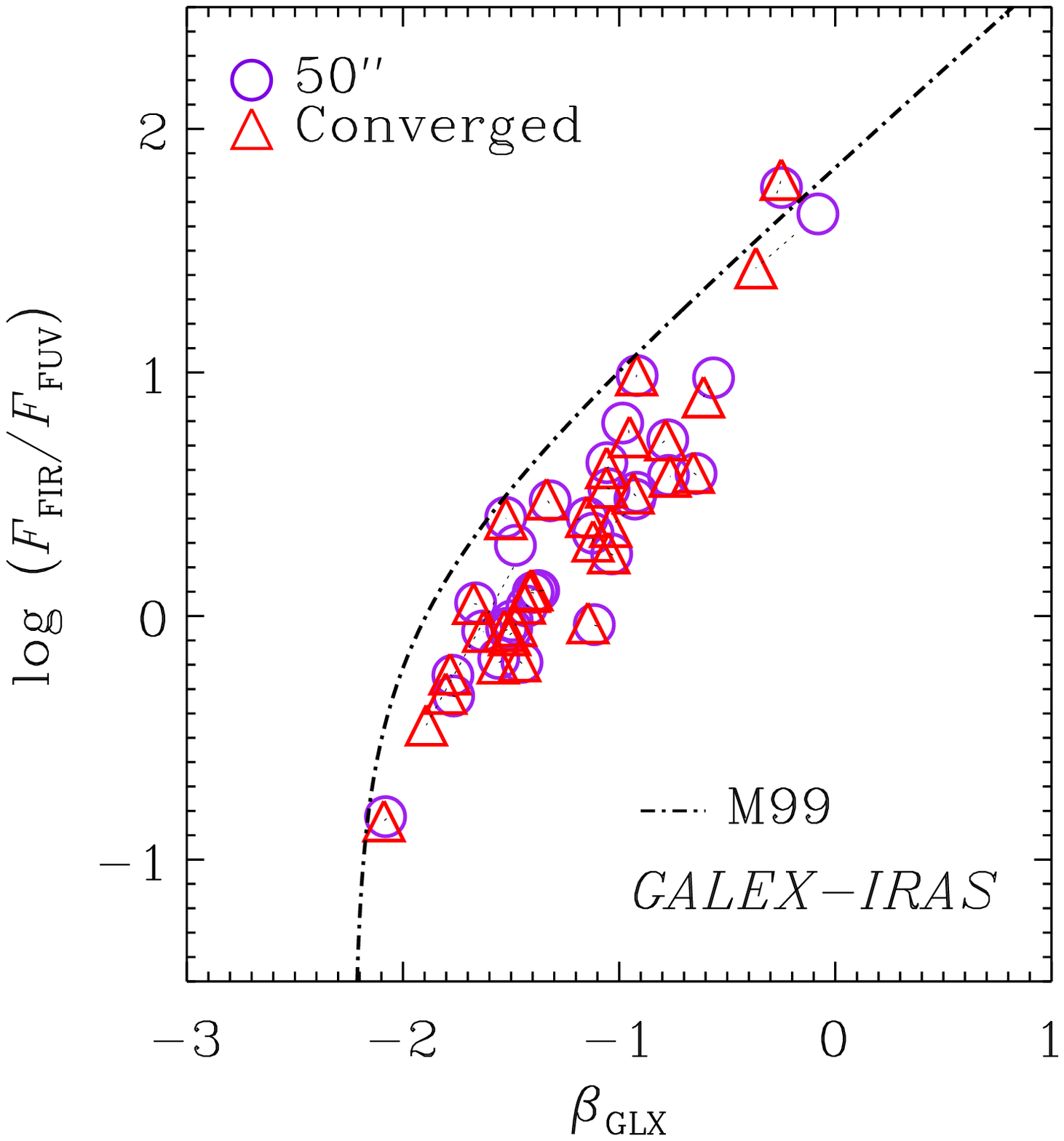}
\caption{
Effect of aperture size for the IRX-$\beta$ measurement. 
These figures are essentially the same as Fig.~\ref{fig:IRX_beta_all}, but we show the change
of the measured values with increasing aperture size with a step of $10''$ here. 
Upper-left: change of the IRX-$\beta$ relation with an aperture radius $10'' \rightarrow 20''$; 
Upper-right: same but radius $20'' \rightarrow 30''$; 
Middle-left: same but radius $30'' \rightarrow 40''$; 
Middle-right: same but radius $40'' \rightarrow 50''$; 
Lower:  same but radius $50'' \rightarrow \mbox{converged values}$.
}\label{fig:five_aperture}
\end{figure*}

\subsection{Result with {\galex} and {\iras}}\label{subsec:galex_iras}

We show the result of new measurements for M99's sample galaxies to
examine the IRX-$\beta$ relation.
We quantitatively define the slope of UV spectrum $\beta$ as the exponent of 
an approximated spectrum (in flux density per unit wavelength interval)
\begin{equation}
 f_{\lambda}\propto\lambda^{\beta} \; .
\end{equation}
The relation between the flux at wavelength $\lambda$, $F_\lambda$, and the flux density per 
unit wavelength interval at $\lambda$, $f_\lambda$ is 
\begin{eqnarray}\label{eq:flux_flux_density}
  F_{\lambda}=\lambda f_{\lambda} = \nu f_\nu \; .
\end{eqnarray}
Thus the 1600~\AA\ flux $F_{1600}$ is defined as the average flux around the central
wavelength 1600~\AA , 
\begin{equation}\label{eq:f1600}
 F_{1600} \equiv \lambda f_\lambda \mbox{@1600~\AA} \; .
\end{equation}
The FIR flux $F_{\rm FIR}$ is defined by using flux densities $f_\nu$ at \iras\ $60~\mu $m 
and $100~\mu$m as
\begin{eqnarray}\label{eq:helou}
  &&F(60) = 2.58 \times 10^{-40}f_{\nu}(60~\mu \rm{m}) \;,\\
  &&F(100) = 1.00 \times 10^{-40}f_{\nu}(100~\mu \rm{m})\;, \\
  &&F_{\rm FIR} = 1.26[F(60)+F(100)] \;, 
\end{eqnarray}
where the unit of $F(60)$, $F(100)$, and $F_{\rm FIR}$ is $[\rm{erg~s^{-1}~cm^{-2}}]$, 
and the unit of $f_{\nu}$ is $~[\rm{erg~s^{-1}~cm^{-2}~}~Hz^{-1}]$ \citep{helou88}.
In this work, we denote the flux density per unit frequency interval at a wavelength $\lambda$
as $f_\nu (\lambda)$.
We also use flux density in unit wavelength interval, $f_\lambda$.
The unit of $f_\lambda$ is $\; [\mbox{erg}^{-1} \mbox{cm}^{-2} \mbox{s}^{-1} 
\mbox{\AA}^{-1}]$ throughout this paper.
In M99\nocite{meurer99}, the IR and the UV fluxes were estimated from the \iras\ and
{\iue} data, respectively. 
Hence, we first use the original \iras\ values of M99 for IR flux, 
and focus on the new UV measurements with {\it GALEX}.  
We should note that the FIR flux evaluated by eq.~(\ref{eq:helou}) is only the integrated 
flux over a wavelength range $42\mbox{--}122~\mu$m, and not the total IR flux
often defined as integrated flux in a range $8\mbox{--}1000\;\mu\mbox{m}$.
We evaluate the IR flux with \akari\ later in Section~\ref{subsec:akari}.

With these observables and some empirical assumptions, M99\nocite{meurer99} have 
formulated the relation between $\mbox{IRX}_{1600}$ and $A_{1600}$ as
\begin{eqnarray}\label{eq:irx_extinction_general}
  &&{\rm log(IRX_{1600})} = {\rm log} (10^{0.4A_{1600}}-1)+\log B \; , \\
  &&B \equiv \frac{\rm {BC(1600)_*}}{\rm BC(FIR)_{Dust}} = \mbox{const.} \;,
\end{eqnarray}
where BC stands for the bolometric correction. 
M99\nocite{meurer99} adopted $\rm{BC}(1600)_*=1.66\pm 0.15$ for 1600~\AA\ and
BC(FIR)$_{\rm Dust}=1.4\pm 0.2$ for \iras\ FIR, which leads $B=1.19\pm0.20$. 
{We revisit this issue in Section~\ref{subsec:dust_BC}.}
Thus, we obtain 
\begin{eqnarray}\label{eq:irx_extinction}
  {\rm log(IRX_{1600})} = {\rm log}(10^{0.4A_{1600}}-1)+0.076 \pm 0.044 \;.
\end{eqnarray}
The derivation of eqs.~(\ref{eq:irx_extinction_general}) and 
(\ref{eq:irx_extinction}) is shown in Appendix~\ref{sec:meurer}.
Here, the value $B$ is for {\iue} and {\iras}, and if we use different data, 
it will also change.
Since we use \galex\ FUV, we should recalculate $B$ for \galex\ and \iras.
However, \citet{seibert05} obtained $\rm{BC}(\mbox{FUV})_* \simeq 1.68$, 
which is very close to that of {\iue}, yielding practically the same value for $B$.
Assuming a linear relation between $A_{1600}$ and $\beta$ as
\begin{eqnarray}\label{eq:ext_beta}
  A_{1600} = a_0 + a_1 \beta \;, 
\end{eqnarray}
namely the general relation between IRX and $\beta$ is expressed as
\begin{equation}\label{eq:irx_beta_meurer_general}
  \log \left( \frac{F_{\rm FIR}}{F_{\rm FUV}}\right)=\log \left[ 10^{(0.4 (a_0+a_1\beta ) }-1 \right] +0.076 \;.
\end{equation}
Performing a least square fit with observed values, M99 obtained a result 
\begin{equation}\label{eq:ext_beta_meurer}
A_{1600}=4.43+1.99\beta \; .
\end{equation}
Combining eqs.~(\ref{eq:ext_beta_meurer}) with (\ref{eq:irx_beta_meurer_general}), we obtain
\begin{equation}\label{eq:irx_beta_meurer}
  \log \left( \frac{F_{\rm FIR}}{F_{\rm FUV}}\right)=\log \left[ 10^{0.4( 4.43 + 1.99\beta )}-1 \right] +0.076 \;.
\end{equation}
This is the IRX-$\beta$ relation proposed by M99\nocite{meurer99}.

\galex\ is a survey-oriented astronomical satellite, and its data are mainly images (though
\galex\ is also equipped with a spectrograph). 
Hence, we need to determine the UV spectral slope $\beta$ from FUV and NUV images,
different from the case of {\iue} which is equipped with a spectrograph.
\citet{kong04} proposed a method to estimate IRX and $\beta$ by using a flux density per
unit wavelength $f_{\rm FUV}$ and $f_{\rm NUV}$ as follows. 
\begin{eqnarray}
  \mathrm{IRX} &\equiv& \log\left(\frac{F_{\rm FIR}}{F_{\rm FUV}} \right) \label{eq:IRX_galex}\\ 
  \beta_{{\rm GLX}} &\equiv& \frac{\log f_{\rm FUV}-\log f_{\rm NUV}}{\log \lambda_{\rm FUV}-\log \lambda_{\rm NUV}}
  \label{eq:beta_galex} \;. 
\end{eqnarray}
Here the effective wavelength of FUV and NUV bands are $\lambda_{\rm FUV}=1520$~\AA\ and
$\lambda_{\rm NUV}=2310$~\AA, respectively.

The IRX-$\beta$ relation newly obtained by these quantities is shown in Fig.~\ref{fig:irx_beta_galex_iras}. 
We also show the original values of M99\nocite{meurer99} for comparison.
Same galaxies are connected with dotted lines in Fig.~\ref{fig:irx_beta_galex_iras}.
The new values distribute clearly below the original ones of M99
for almost all galaxies.
We found that since also the new values for $\beta$ are different, the distribution is shifted
toward lower-right direction. 
Fitting eq.~(\ref{eq:irx_beta_meurer_general}) to the new {\galex}-{\iras} IRX-$\beta$ measurements, 
we obtain 
\begin{equation}\label{eq:irx_beta_galex_iras}
  \log \left( \frac{F_{\rm FIR}}{F_{\rm FUV}}\right) = \log \left[ 10^{(0.4 (3.45 + 1.82\beta ) } -1 \right] +0.076 \;,
\end{equation}
which is different from eq.~(\ref{eq:irx_beta_meurer}) (see the 
dashed line in Fig.~\ref{fig:irx_beta_galex_iras}).
{ 
The newly obtained relation is very close to that of normal star-forming galaxies proposed
by previous studies \citep[e.g., ][, among others]{buat05,cortese06,boissier07,takeuchi10}.
This implies that the starbursts in M99 sample are hosted in a normal galaxies with redder
colors.
We will revisit this issue in Section~\ref{sec:discussion}.
}

\begin{figure*}[tb]
\centering\includegraphics[width=0.8\textwidth]{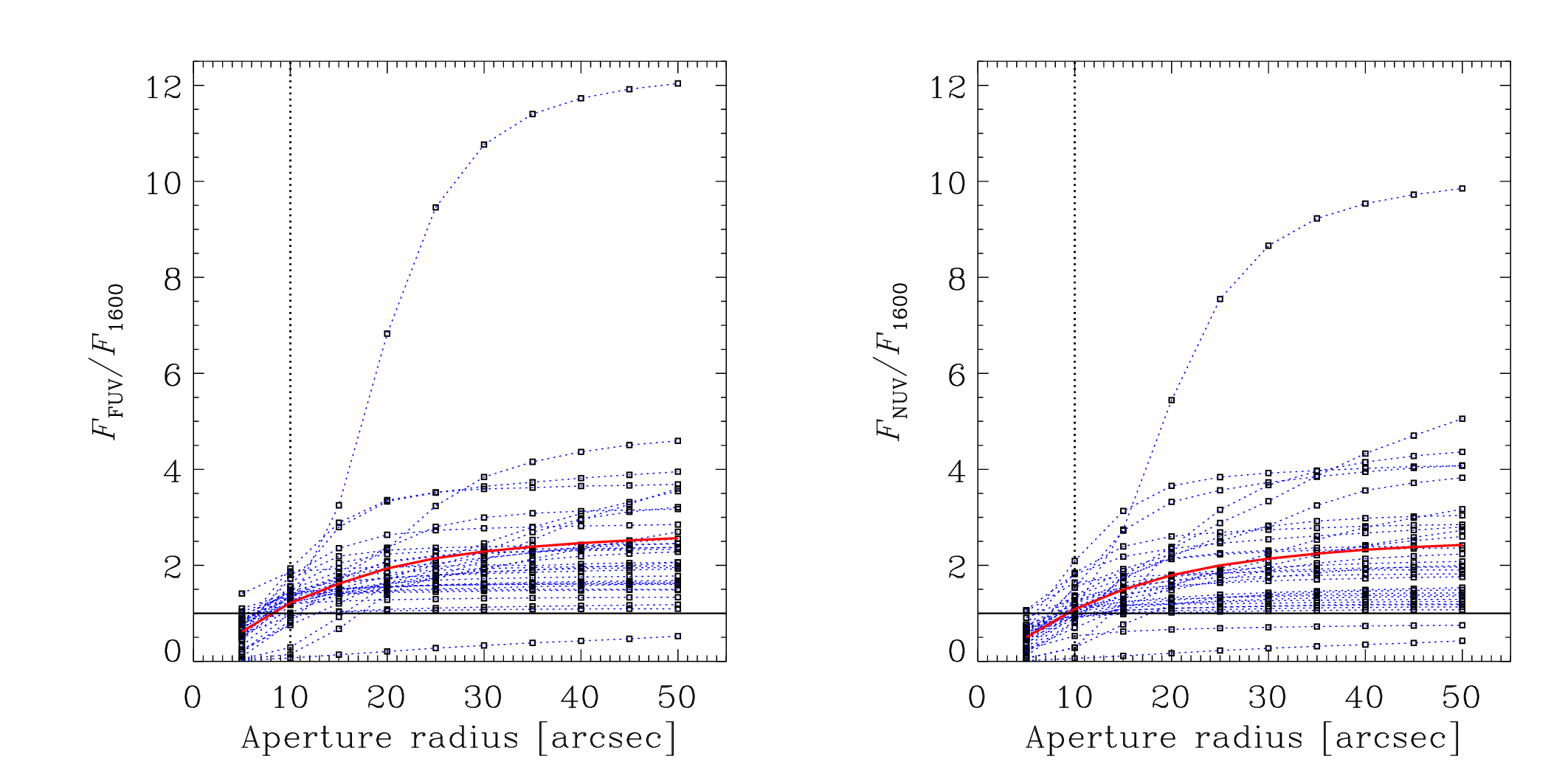}
\caption{
Behavior of {\galex} FUV and NUV fluxes normalized by the {\iue}~1600~\AA\ flux. 
Left: the flux ratio between {\galex}~FUV and {\iue}~1600~\AA.
Open squares connected by dotted lines represent the flux ratio for each sample 
galaxy as a function of the aperture radius.
Solid thick curve is the average of all sample galaxies;
Right: same as left panel but for {\galex}~NUV.
{ A radius $10''$ is the semi-major axis radius of \iue\ aperture (see main text), 
which is indicated by vertical dotted lines in both panels.
}
}\label{fig:frac_galex_1600}
\end{figure*}

\begin{figure*}[tb]
   \centering\includegraphics[width=0.7\textwidth ]{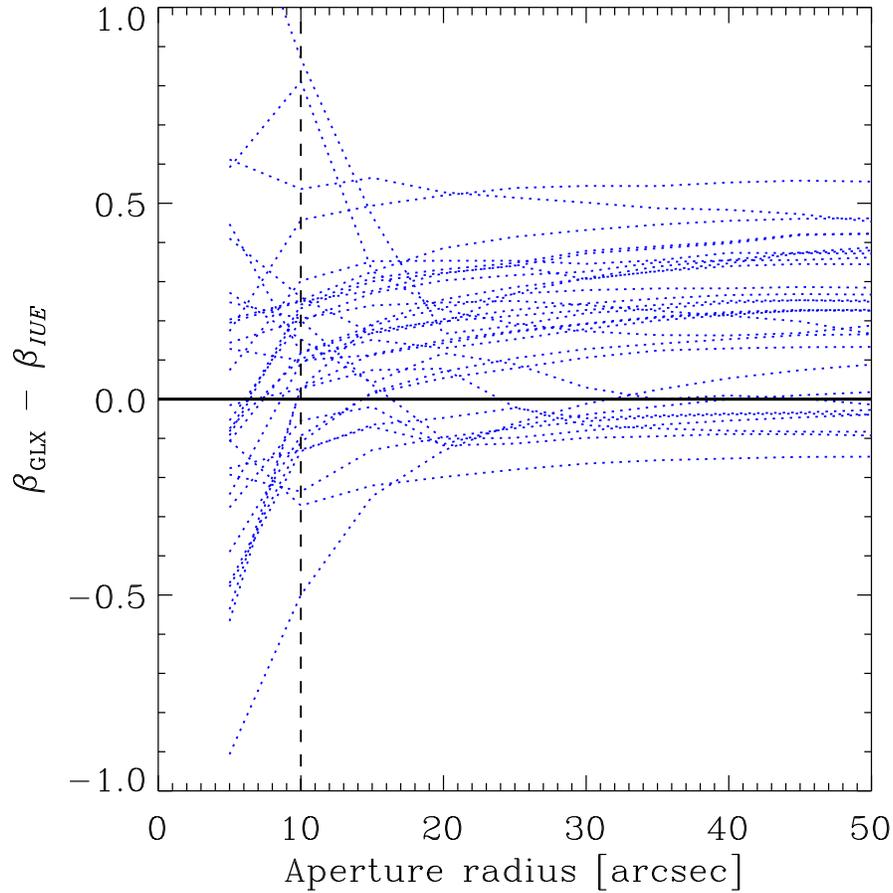}
\caption{
Behavior of the difference between $\beta_{{\rm GLX}}$ and $\beta_{IUE}$
as a function of aperture radius of our \galex\ photometry.
The slope from {\iue}, $\beta_{IUE}$ is fixed to the original M99 value for
each sample galaxy. 
A radius $10''$ is the semi-major axis radius of \iue\ aperture (see main text), 
which is indicated by vertical dashed line.
The horizontal solid line means $\beta_{\rm GLX} = \beta_{IUE}$.
{ Again the radius $10''$ is indicated by vertical dashed line.
}
}\label{fig:beta_radius}
\end{figure*}

\subsection{Examination of the result}

\begin{figure*}[tb]
\centering\includegraphics[width=0.9\textwidth]{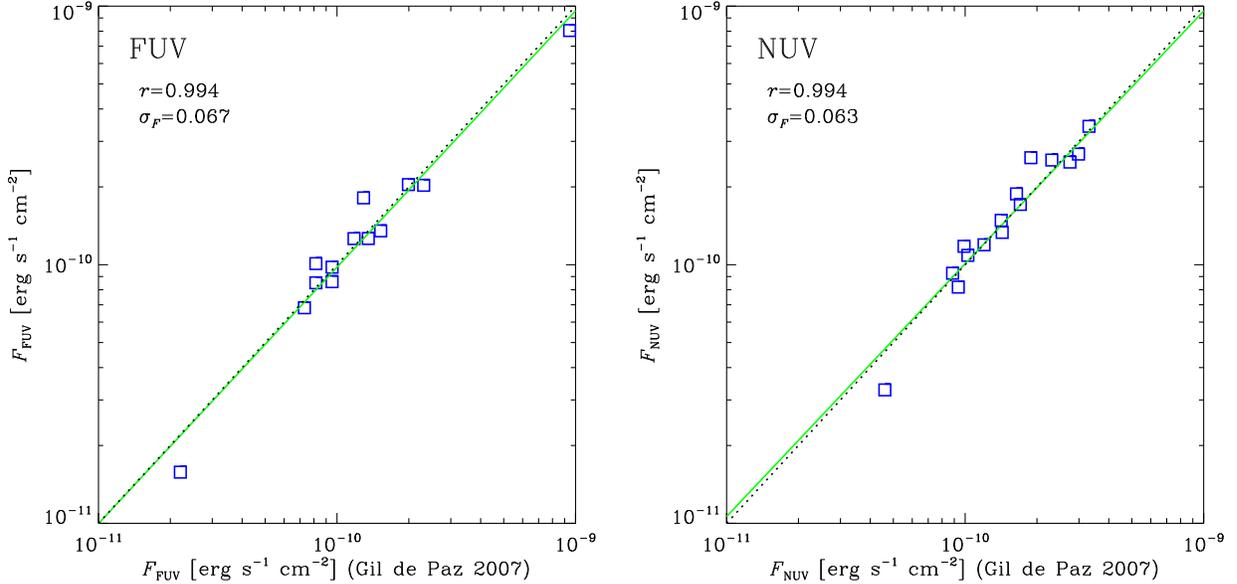}
\caption{
Comparison between the measurements of this work and that of 
\citet{gil_de_paz07}.
Left panel shows the correlation of two measurements at FUV, and right panel
is the same but at NUV.
In both panels, open squares represent 11 galaxies commonly included both in the sample of
\citet{gil_de_paz07} and of this work.
Solid lines represent the logarithmic linear fit to the sample. 
Dotted diagonal lines show $y = x$.
}\label{fig:gil_de_paz}
\end{figure*}

\subsubsection{Effect of the filter response function of {\galex}}\label{subsubsec:filter}

As we mentioned above, \galex\ data are basically images, then the photometric data
would be affected by the filter response function.
We should stress that the UV spectral slopes $\beta$ were obtained directly from
the \iue\ spectrum, with avoiding the wavelength range corresponding to the 
2170~\AA\ bump \citep[see e.g.][]{cardelli89} of the Galactic extinction curve \citep{calzetti94}.
In contrast, the \galex\ NUV filter covers the bump.
This would cause a systematic effect on the measurement of IRX and $\beta$.
Since most of the high-$z$ galaxy surveys use photometric data to evaluate $\beta$ for 
applying the IRX-$\beta$ relation fort dust attenuation, it is crucial to calibrate the related 
quantities carefully.
We first examine this effect.

For this, we convolve the response function with the \iue\ energy spectra of the M99 sample galaxies
(see Fig.~\ref{fig:NGC4861} as an example).
Let $R_{\rm FUV}(\lambda)$ and $R_{\rm NUV}(\lambda)$ be the FUV and NUV filter 
response functions, respectively, and denote the raw \iue\ spectra as $f_{\lambda}$ and the flux density at
FUV (NUV) as $f_{\rm FUV,fil}$ ($f_{\rm NUV,fil}$) $[\mbox{erg}^{-1}\mbox{s}^{-1}\mbox{cm}^{-2}\mbox{\AA}^{-1}]$, 
then we have 
\begin{eqnarray}
 f_{\rm FUV,\rm{fil}} &=& \frac{\displaystyle\int_{\rm FUV}f_{\lambda}R_{\rm FUV}(\lambda )d\lambda}{\displaystyle
 \int_{\rm FUV}R_{\rm FUV}(\lambda )d\lambda} \label{eq:flux_fil_galex_fuv}\\
 f_{\rm NUV,\rm{fil}} &=& \frac{\displaystyle\int_{\rm NUV}f_{\lambda}R_{\rm NUV}(\lambda )d\lambda}{\displaystyle
 \int_{\rm NUV}R_{\rm NUV}(\lambda )d\lambda} \label{eq:flux_fil_galex_nuv} \;.
\end{eqnarray}
With eqs.~(\ref{eq:flux_fil_galex_fuv}) and (\ref{eq:flux_fil_galex_nuv}), IRX and $\beta$ after transmitting 
the \galex\ filters, $\mbox{IRX}_{fil}$ and $\beta_{,\rm{fil}}$, are expressed as
\begin{eqnarray}
  &&\mathrm{IRX_{\rm{fil}}} = \log\left(\frac{ F_{\rm FIR}}{F_{\rm FUV,\rm{fil}}} \right) \label{eq:IRX_fil_galex}\\ 
  &&\beta_{{\rm GLX}_{\rm{fil}}} = 
    \frac{\log f_{\rm FUV,\rm{fil}}-\log f_{\rm NUV,\rm{fil}}}{\log \lambda_{\rm FUV}-\log \lambda_{\rm NUV}}\\
  &&F_{\rm FUV,\rm{fil}} = \lambda_{\rm FUV} f_{\rm FUV,\rm{fil}}
	\label{eq:beta_fil_galex} \;.
\end{eqnarray}

We show the comparison of the UV spectral slope $\beta$ and the 1600~\AA\ flux density{\galex} before and
after transmitting the \galex\ filter in Fig.~\ref{fig:iue_galex_filter}. 
{For this analysis, we restricted to galaxies only with spectra in the full range of {\iue} wavelength coverage
($1100\mbox{--}3000$~\AA).
In the M99 sample, there are samples only with spectra at \iue\ SW spectral coverage, and they are not ideal for 
the purpose of this analysis here.}
As we see in the left panel of Fig.~\ref{fig:iue_galex_filter}, $\beta$ directly measured by
\iue\ and that defined by \galex\ show a very good agreement. 
The linear fit gives
\begin{eqnarray}
  \beta_{\iue} = 1.01 \beta_{{\rm GLX, fil}} -0.07
\end{eqnarray}
with the Pearson's correlation coefficient of $r = 0.976$ and
the standard deviation is $\sigma_\beta \simeq 0.14$.
The systematic deviation is smaller than $\sigma_\beta$.
As for the flux density, as we see on the right panel in Fig.~\ref{fig:iue_galex_filter}, 
the 1600~\AA\ flux density measured by {\iue}, $F_{1600}$, tends to be about 10~\% larger
than that transmitted through the \galex\ FUV filter, $F_{\rm FUV,fil}$.
The logarithmic linear fit yields
\begin{eqnarray}
  \log F_{1600} = 0.949 \log F_{\rm FUV, fil} -0.462
\end{eqnarray}
with the Pearson's correlation coefficient of $r = 0.950$ and
the standard deviation of flux density $\sigma_F \simeq 0.11$
in logarithmic scale.
Again, the systematic deviation is smaller than $\sigma_F$.
Considering the intrinsic scatter in the IRX-$\beta$ plot in M99, we can 
safely neglect the systematic differences in $\beta$ and $F_{\rm FUV\; or\; NUV}$
in the following analysis.

\subsubsection{Effect of the aperture size}

In order to examine the effect of aperture photometry on \galex\ images, we observe the
behavior of flux with varying the aperture size.
In this work, the aperture radius $\theta''$ of the photometry is defined as an ellipse with
semi-major axis size $\theta''$.  
This ellipse is determined automatically by fitting to the isophotal ellipse of a galaxy
by our photometry software.
Then, the flux within a radius of $10''$ means the integrated flux in an ellipse with
semi-major axis size of $10''$.
Figure~\ref{fig:IRX_beta_all} shows the change of measured values on the IRX--$\beta$ plot
when we change the {\galex} aperture radius as $5''$, $10''$, $15''$, $20''$, and the
convergence radius (depends on the sample). 
{}To see the change more clearly, we also show the change
of the measured values with increasing aperture size with a step of $10''$ in Fig.~\ref{fig:five_aperture}. 

In Fig.~\ref{fig:frac_galex_1600}, we compare {\galex} FUV and NUV flux densities and {\iue}~$F1600$ flux. 
{Here, the circle-equivalent ``effective'' \iue\ aperture of $14''$ is shown as vertical lines.
This is an approximate of the \iue\ aperture size which was an ellipse of $10''\times20''$.
We set the ellipse shape and position angle to be the same as that used in actual \iue\ observation 
of the sample galaxies.
This \iue\ effective aperture is depicted by vertical lines in both panels of Fig.~\ref{fig:frac_galex_1600}.
As shown in this figure, FUV fluxes do not start to converge around a radius of $10''$--$20''$ but 
continue to increase, and finally converge to values approximately 2.5 times larger than the 
1600~\AA\ fluxes.
The NUV flux also show a similar behavior to that of FUV.
We should also note that the behavior of FUV and NUV growth curves is
not exactly the same.}
This causes an aperture dependence of $\beta$.

{}To see this effect, we examined the difference of $\beta$ between \iue\ and \galex.
Figure~\ref{fig:beta_radius} depicts the behavior of $\beta_{\rm GLX} - \beta_{IUE}$ as a function of 
aperture radius.
We observe that the behavior of $\beta$ is not monotonic: some increase and others 
decrease after exceeding $14''$ aperture, and finally start to converge to certain values at much
larger radius.
Within the effective \iue\ aperture radius, $\beta_{{\rm GLX}}$ changes  
significantly for most of the sample and is still far from the convergence values.
We see a weak tendency that the converged values of $\beta$ are larger than that listed in M99,
i.e., the global UV color of the sample turned out to be slightly redder when we measure the integrated
flux.
This can be explained as follows: since the M99 sample consists of central starbursts, it may be 
natural that their outer part of the disk has redder colors than the central part.
Hence, the colors of central part are bluer than the global ones.
{Thus, the values of $\beta$ determined at a radius similar to that of \iue\ are systematically
bluer than the UV slopes when larger apertures are used.

We, therefore, conclude that the systematic deviation of the 
IRX-$\beta$ relation from the original one is due to the effect of UV flux 
because of the small \iue\ aperture.}
Especially, we see there is a galaxy that shows a very prominent behavior
in Fig.~\ref{fig:frac_galex_1600}.
Both at FUV and NUV, this galaxy converges to exceptional values much brighter than
the fluxes at a radius of $10''$.
This galaxy is NGC~2537, which has very strong contribution of UV flux from the
peripheral regions.
Because of this, for NGC~2537, the \iue\ flux is more than an order of magnitude 
underestimated.

\subsubsection{Verification of the result with \citet{gil_de_paz07}}

\begin{figure*}[t]
\centering\includegraphics[width=0.7\textwidth]{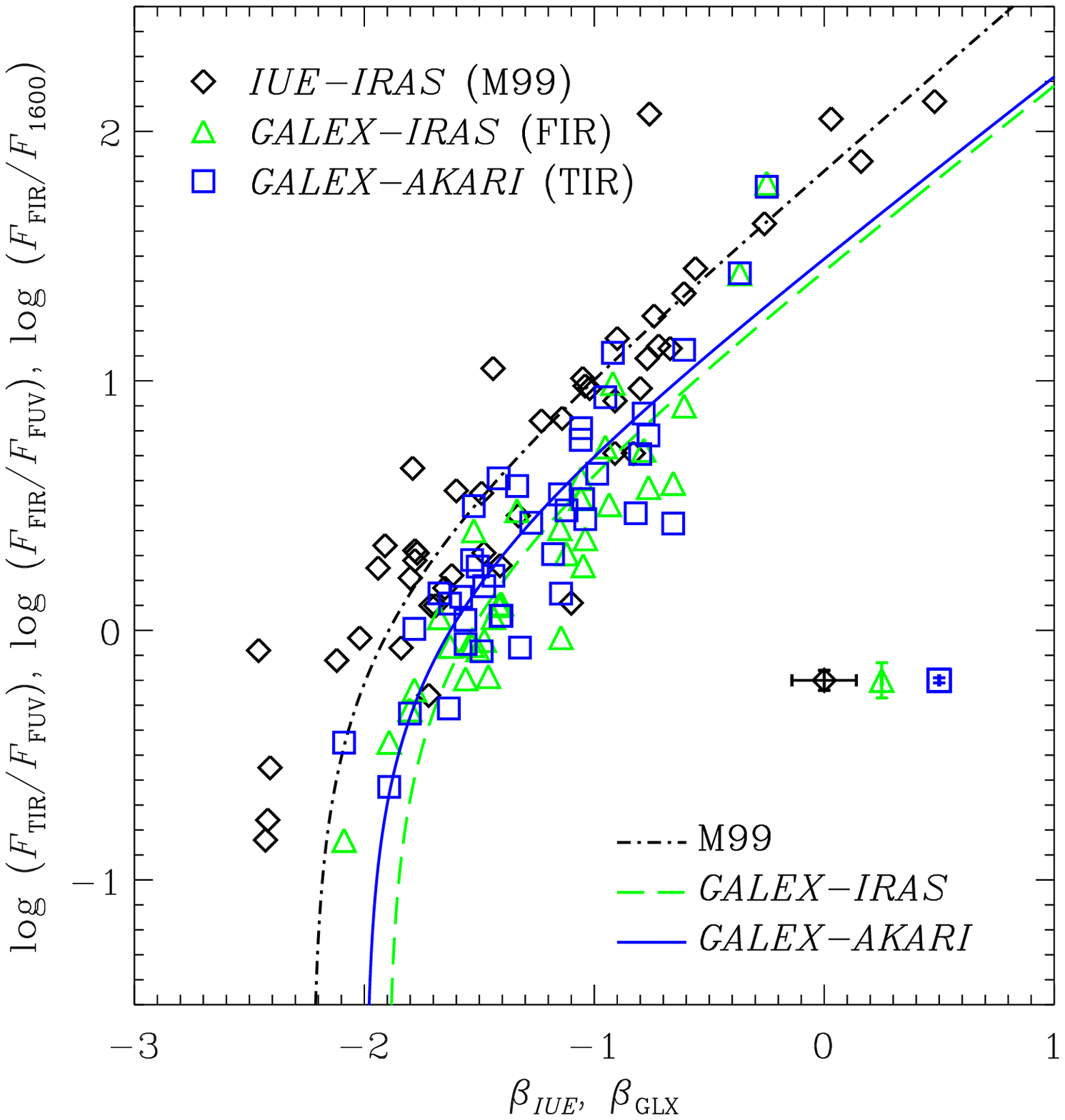}
\caption{
New IRX-$\beta$ relation obtained by \galex\ and \akari\ diffuse map.
Measurements with {\iue}-{\iras}, {\galex}-{\iras}, and {\galex}-{\akari} are plotted.
Diamonds, triangles, and squares represent the {\iue}-{\iras}, {\galex}-{\iras}, and {\galex}-{\akari}
values.
Dot-dashed curve represents the original M99 formula [eq.~(\ref{eq:irx_beta_meurer})], 
dashed curve for {\galex} -{\iras} fit [eq.~(\ref{eq:irx_beta_galex_iras})],  
and solid curve  for the new formula with {\galex}-{\akari} measurements [eq.~(\ref{eq:irx_beta_akari})].
Note that for {\iue}-{\iras} and {\galex}-{\iras}, the IR flux is evaluated by FIR \citep{helou88},
while for {\galex}-{\akari} it is evaluated by total IR emission \citep[TIR: ][]{takeuchi10}.
}\label{fig:irx_akari_galex}
\end{figure*}

In the previous sections, we have shown that the UV flux measurement of M99 was significantly
underestimated because of the small aperture size of {\iue}.
However, since the photometry software we used for this work has been developed originally 
by ourselves \citep[cf.][]{iglesias06}, we should examine if there would be any systematic effect 
caused by the difference in photometry softwares when we compare our results with other ones.

\citet{gil_de_paz07} also used original software to perform a photometry of {\galex} images.
We checked the sample of \citet{gil_de_paz07} and found 18 galaxies common with M99, among
which we have 11 galaxies with measurements needed for this study \citep[][see their Table~4]{gil_de_paz07}. 

Figure~\ref{fig:gil_de_paz} shows that the results of this work and that of \citet{gil_de_paz07} agree
with each other very well.
{The correlation coefficient and the standard deviation for FUV data are $r = 0.994$ 
and $\sigma_F({\rm FUV}) = 0.0670$ in logarithmic scale, and the same for NUV data 
are $r = 0.994$ and $\sigma_F({\rm NUV}) = 0.0630$, respectively. 
Thus, we can safely conclude that both measurements are in excellent agreement.
}
This means that the aperture effect on the original IRX-$\beta$ relation was independently 
proved by \citet{gil_de_paz07}.
Now we can safely conclude that the reason of the systematic shift of the IRX-$\beta$ relation
of recent studies and the original one is dominantly caused by the small aperture of \iue.
Now we have the {\galex}-{\iras} based IRX-$\beta$ relation which is free from the aperture effect 
[eq.~(\ref{eq:irx_beta_galex_iras})].

\subsection{New IRX-$\beta$ relation with {\galex} and {\akari}}

Up to now, we concentrated on the examination of UV flux measurement for the IRX-$\beta$
relation, and for this purpose we kept the M99's original values $F_{\rm FIR}$ from \iras.  
Here we discuss the IR flux. 

\subsubsection{Total IR flux by \akari}

The M99 relation was constructed to relate the UV slope and IRX $(F_{\rm FIR}/F_{\rm 1600})$,
and they used the FIR luminosity \citep{helou88}
\begin{equation}\label{eq:fir_helou}
 L_{\rm FIR}\equiv 3.29\times 10^{-22}\times \left[ 2.58L_{\nu}(60\;\mu \mbox{m})+L_{\nu}(100\;\mu \mbox{m})
 \right] \; [\rm{L}_{\odot}] \; .
\end{equation}
However, as mentioned above, $L_{\rm FIR}$ contains only the luminosity in a wavelength 
range of $\lambda =42-122~\mu $m, and energy emitted at MIR ($5\mbox{--}30~\mu$m) nor submillimeter 
($300\mbox{--}1000~\mu$m) is not considered.
Hence, the FIR luminosity is proved to be a significant underestimation as a representative of the total IR luminosity ($L_{\rm TIR}$) 
\citep[e.g.,][]{takeuchi05b}.
\citet{dale01} and \citet{dale02} proposed formulae to convert FIR into TIR luminosity 
by making use of \iras\ flux densities. 

The \akari\ FIS All-Sky Survey data are publicly available.
With \akari\ fluxes, \citet{takeuchi10} proposed a new simple formula to estimate $L_{\rm TIR}$ 
precisely as follows\footnote{\citet{hirashita08} constructed a formula to
estimate $L_{\rm TIR}$ by making use of {\it N60}, {\it WIDE-S}, and {\it WIDE-L} flux densities.
However, since {\it N60} flux density is more noisy than the other wide bands, \citet{takeuchi10}
found that the formula using two wide bands [{\it WIDE-S} and {\it WIDE-L}: eq.~(\ref{eq:tir_akari})]
gives a smaller scatter. Hence, we use this two-band based formula in this work.}
\begin{eqnarray}\label{eq:tir_akari}
 &&\Delta\nu(\mbox{\it WIDE-S\,}) = 1.47\times 10^{12}~[\rm{Hz}] \;, \label{eq:l_tir_wideS}\\
 &&\Delta\nu(\mbox{\it WIDE-L\,}) = 0.831\times 10^{12}~[\rm{Hz}] \;, \label{eq:l_tir_wideL}\\
 &&\log L_{\it AKARI} ^ {2 \rm{band}} = \Delta\nu (\mbox{\it WIDE-S\,})
 L_{\nu}(90~\mu \mbox{m}) \;, \nonumber \\ 
 &&\qquad\qquad\qquad+\Delta\nu (\mbox{\it WIDE-L\,})L_{\nu}(140~\mu \mbox{m}) \label{eq:l_tir_2band}\\
 &&\log L_{\rm TIR} = 0.964\log L_{\it AKARI}^{2\rm{band}}+0.814 \label{eq:l_tir} \;.
\end{eqnarray}
Here $\Delta \nu$ is the bandwidth of each \akari\ FIS filter expressed in units of frequency \citep{hirashita08}.

\subsubsection{New IRX-$\beta$ relation with \galex\ and \akari}\label{subsec:akari}

Combining the {\galex}-based UV flux free from the {\iue} aperture effect and
the {\akari} TIR flux [eq.~(\ref{eq:l_tir})] covering the whole IR wavelength range, 
we construct a new IRX-$\beta$ relation applicable for recent studies with \akari , {\it Spitzer}, 
and {\it Herschel}. 
 
We cross-matched the M99 sample with the database of {\akari}~FIS BSC\footnote{DARTS (the {\akari} data archive) 
URL: http://darts.jaxa.jp/astro/akari/cas.html.}.
Among the M99 sample, 48 are detected by {\akari}~FIS.
Among these 48 galaxies, 40 have {\galex}~FUV and NUV (see Table~\ref{table:meurer}).
With these 40 subsample galaxies of M99, we estimated the new IRX-$\beta$ relation. 

Again the UV slope of the SED, $\beta$, was obtained by eq.~(\ref{eq:beta_galex}), 
and the IRX was estimated from {\galex} FUV flux and {\akari} TIR flux as follows. 
\begin{eqnarray}
  \mathrm{IRX} &=& \log\left(\frac{F_{\rm TIR}}{F_{\rm FUV}} \right) \label{eq:IRX_akari} \; . 
\end{eqnarray}
The result is presented in Fig.~\ref{fig:irx_akari_galex}.
Since the {\iras}~FIR is only a part of the {\akari}~TIR, the {\galex}-{\iras} values distribute
in a lower region than the {\galex}-{\akari} measurements on the IRX-$\beta$ plots.

However, we should carefully check one issue here.
Since we tried to be careful about the aperture effect at UV, we should also examine if the
\akari\ BSC fluxes are not affected by the aperture effect.
Importantly, \citet{yuan11} examined a sample of Local galaxies extracted from that of 
\citet{takeuchi10}, since they are mostly nearby resolved galaxies by using \akari\ diffuse map.
They remeasured the IR fluxes with \akari\ diffuse images and found that the flux of extended
galaxies tends to be underestimated in the \akari\ FISBSC, which adopted a complicated 
point source extraction procedure.
Then, according to \citet{yuan11}, we measured the M99 sample with \akari\ FIS diffuse map 
and compared the obtained flux with that in the FISBSC (Appendix~\ref{sec:akari_diffuse}).
Since M99 tried to remove very extended galaxies in the primary construction of the sample,
the current sample was found to be less affected by this effect, but still for some galaxies
the underestimation was significant. 
We thus used the fluxes from the \akari\ diffuse map and used them for the final determination
of the IRX-$\beta$ relation.
This is shown in Fig.~\ref{fig:irx_akari_galex}.

For fitting eq.~(\ref{eq:irx_beta_meurer_general}) to the data, we should recall that this is 
a formula constructed for \iras\ measurement, reflected in the term $\log B$.
Since the bolometric correction for TIR, $\mbox{BC(TIR)}$ is unity by definition, the last
constant term in eq.~(\ref{eq:irx_beta_meurer_general}) [i.e., eq.~(\ref{eq:bc_ratio})] 
simplifies to the logarithm of the bolometric correction at UV, $\log \mbox{BC(FUV)}_* \simeq 0.22$.
Fitting the eq.~(\ref{eq:irx_beta_meurer_general}) to the {\galex}-{\akari} data with this modification
yields the following formula.
\begin{equation}\label{eq:irx_beta_akari}
  \log \left( \frac{L_{\rm TIR}}{L_{\rm FUV}}\right)=\log \left[ 10^{0.4(3.06+1.58\beta )}-1 \right] +0.22 \; .
\end{equation}
This is the new unbiased IRX-$\beta$ relation for UV-selected starburst galaxies free from
any aperture effect in UV or IR, applicable for present-day observational datasets with
UV images (like {\galex}) and total integrated IR flux (like that obtained by {\akari}).

\begin{figure*}[t]
\centering\includegraphics[width=0.7\textwidth]{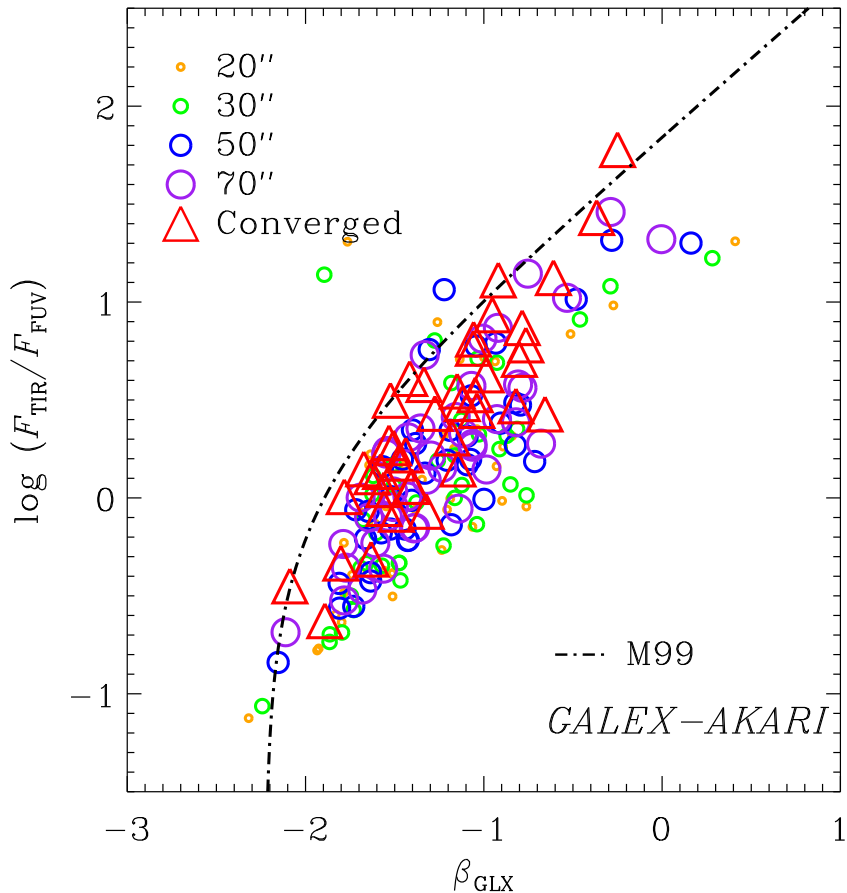}
\caption{
The IRX-$\beta$ diagram as a function of the aperture size both on \galex\
and \akari\ images.
The same elliptical apertures were applied to both of the images.
To address the difference of the angular resolutions of {\galex} and {\akari}, we have convolved
the {\galex} FUV images with the synthesized PSF of {\akari} FIS 140~$\mu$m band. 
Since \akari\ angular resolution is $\sim 50''$ at longer wavelenghts, 
photometric values with aperture radii
smaller than $20''$ are not shown.
Note that the aperture radii are different from that in Fig.~\ref{fig:IRX_beta_all}: 
they are $20''$, $30''$, $50''$, and $70''$.
}\label{fig:galex_akari_2band}
\end{figure*}

\begin{figure*}[p]
\centering\includegraphics[width=0.35\textwidth]{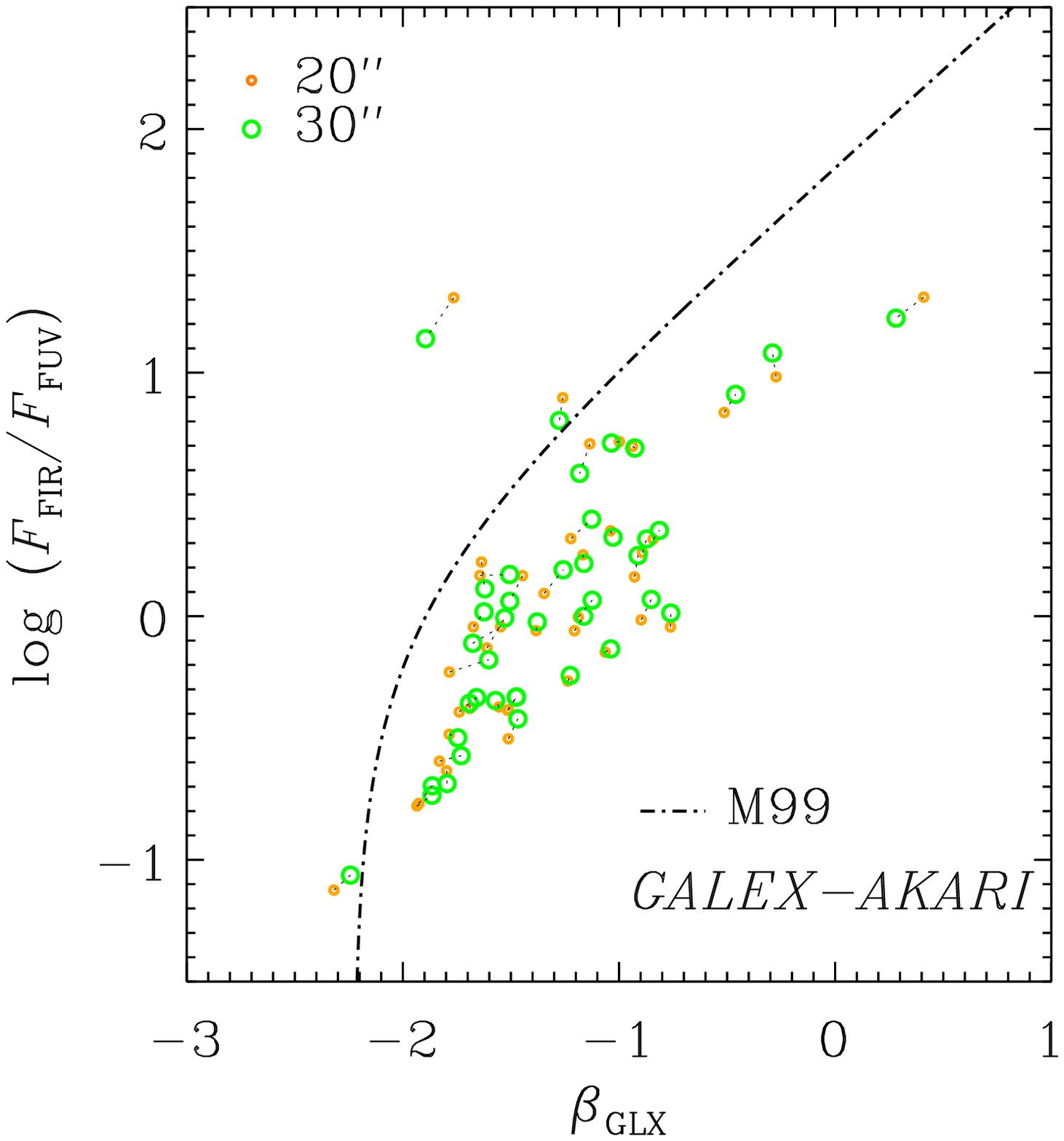}
\centering\includegraphics[width=0.35\textwidth]{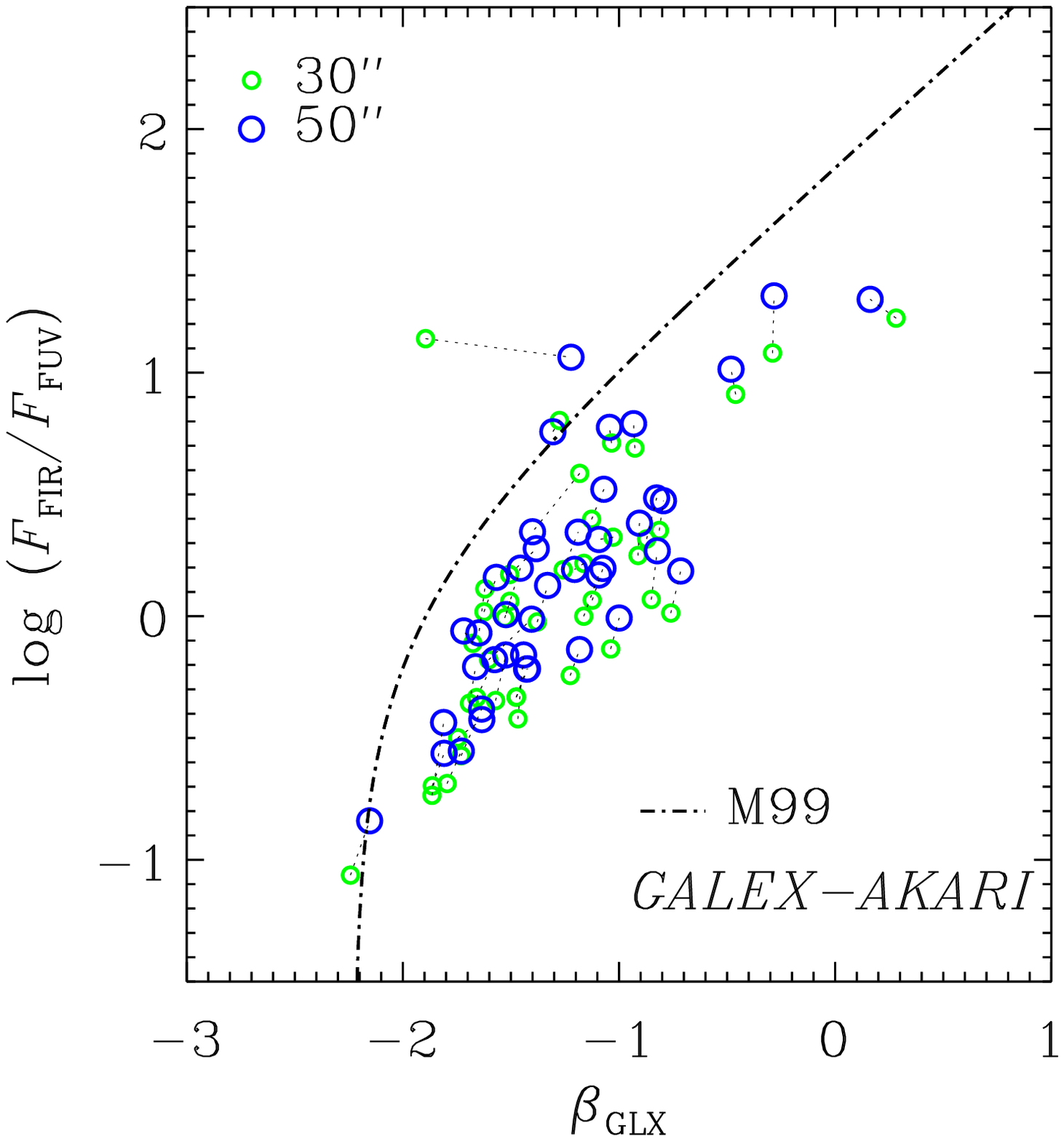}
\centering\includegraphics[width=0.35\textwidth]{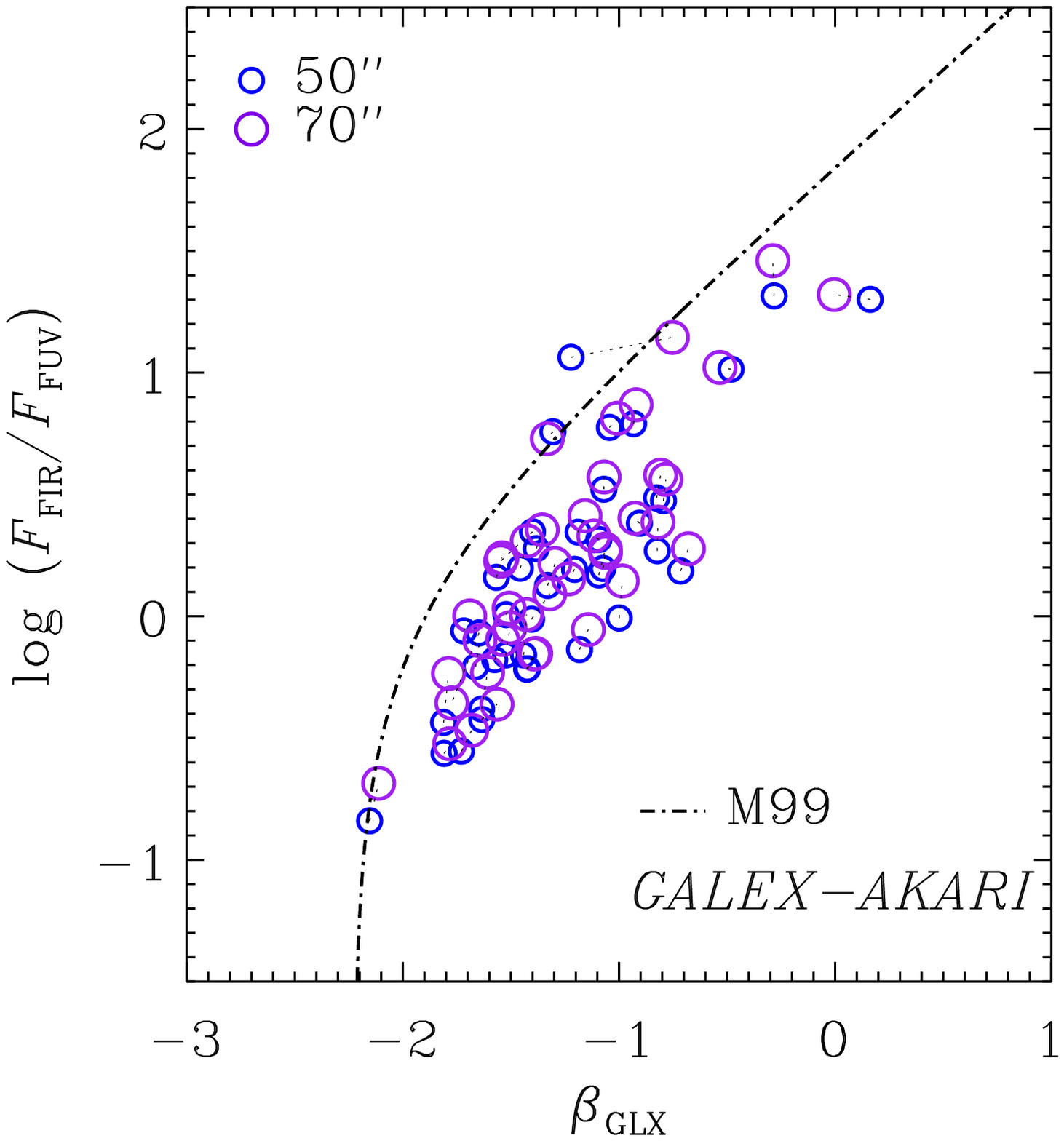}
\centering\includegraphics[width=0.35\textwidth]{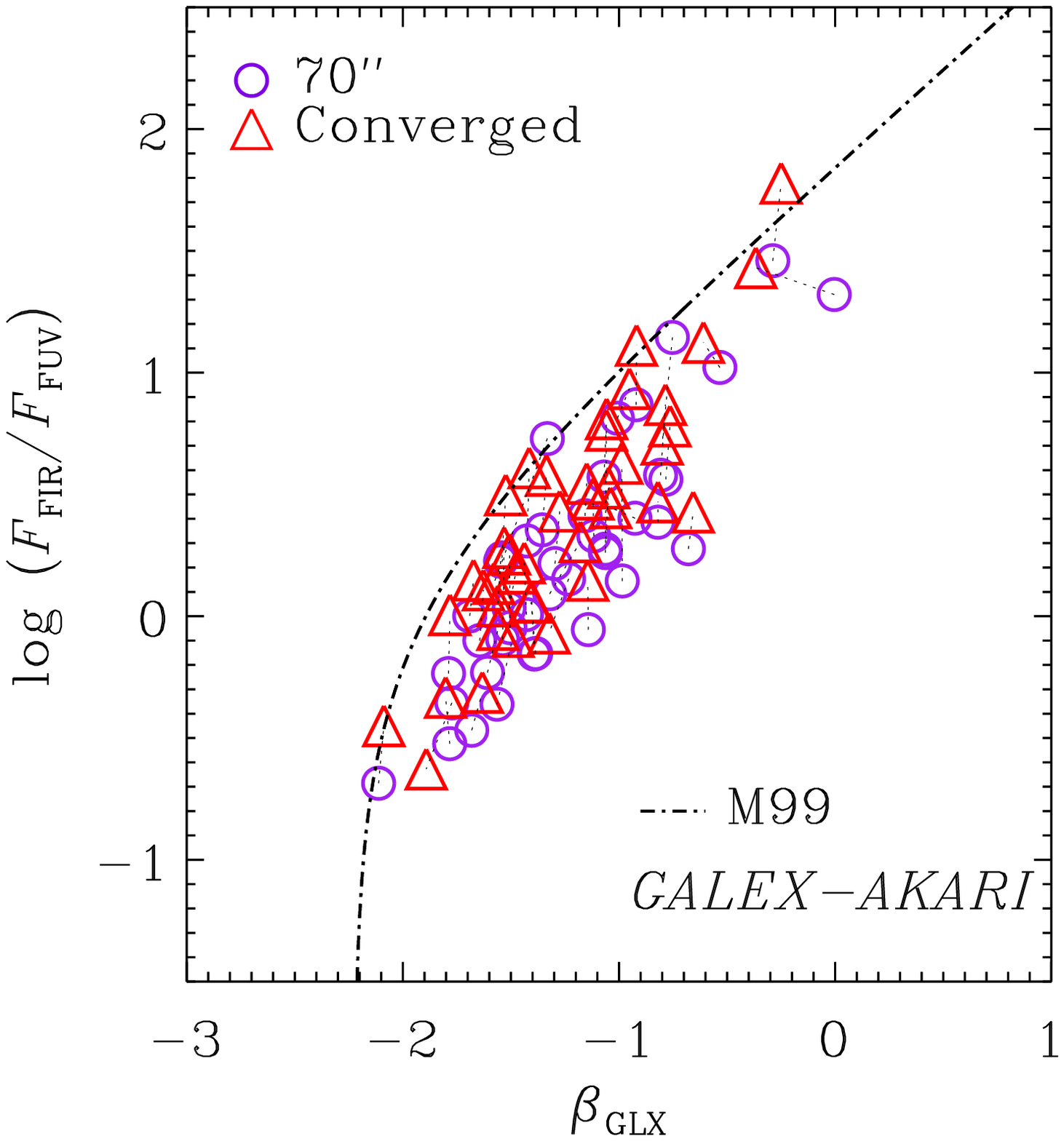}
\caption{
Similar to Fig.~\ref{fig:five_aperture} but with varying aperture size both on \galex\ and \akari\
images. 
{ The {\galex} FUV images are again convolved with the synthesized PSF. 
Since \akari\ angular resolution is $\sim 50''$ at longer wavelenghts, we start the aperture radius from $20''$.
}
}\label{fig:five_aperture_2band}
\end{figure*}

\begin{figure*}[t]
\centering\includegraphics[width=0.7\textwidth]{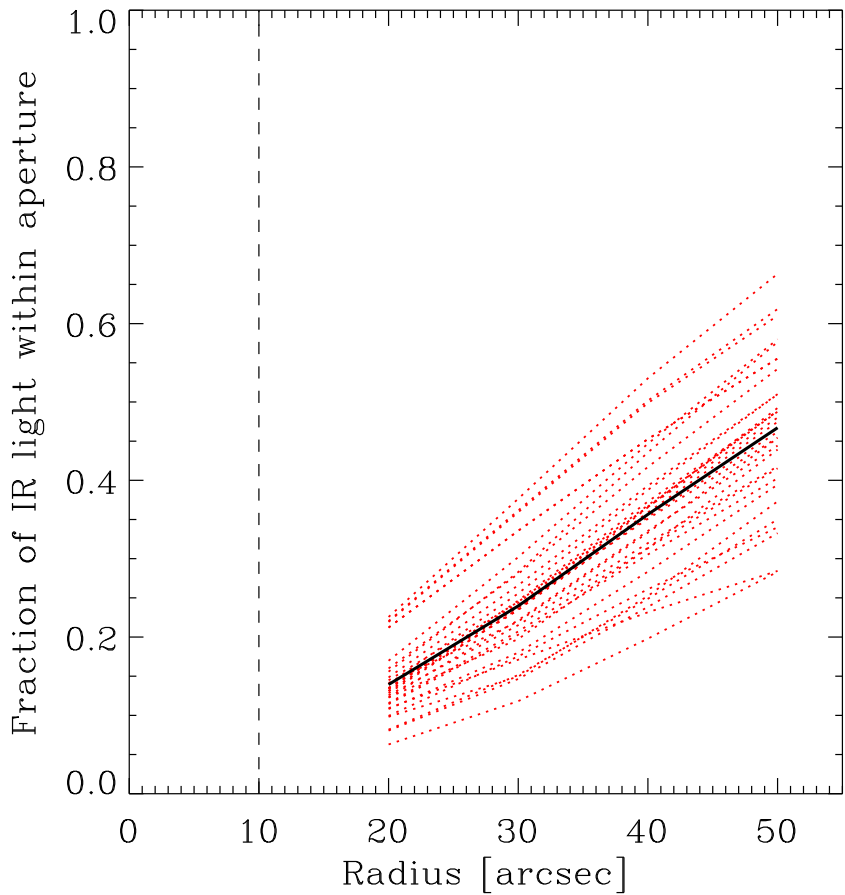}
\caption{
The fraction of the total IR light within an aperture radius.
Dotted lines are the measured fraction for each sample galaxy,
and the thick broken solid line shows their average.
Vertical dashed line represent the semi-major axis of 
\iue\ elliptical aperture.
{ The leak of flux to the sidelobes of the {\akari} PSF is not corrected, 
but this does not change the conclusion significantly.
}
}\label{fig:tir_frac}
\end{figure*}

\begin{figure*}[t]
\centering\includegraphics[width=0.7\textwidth]{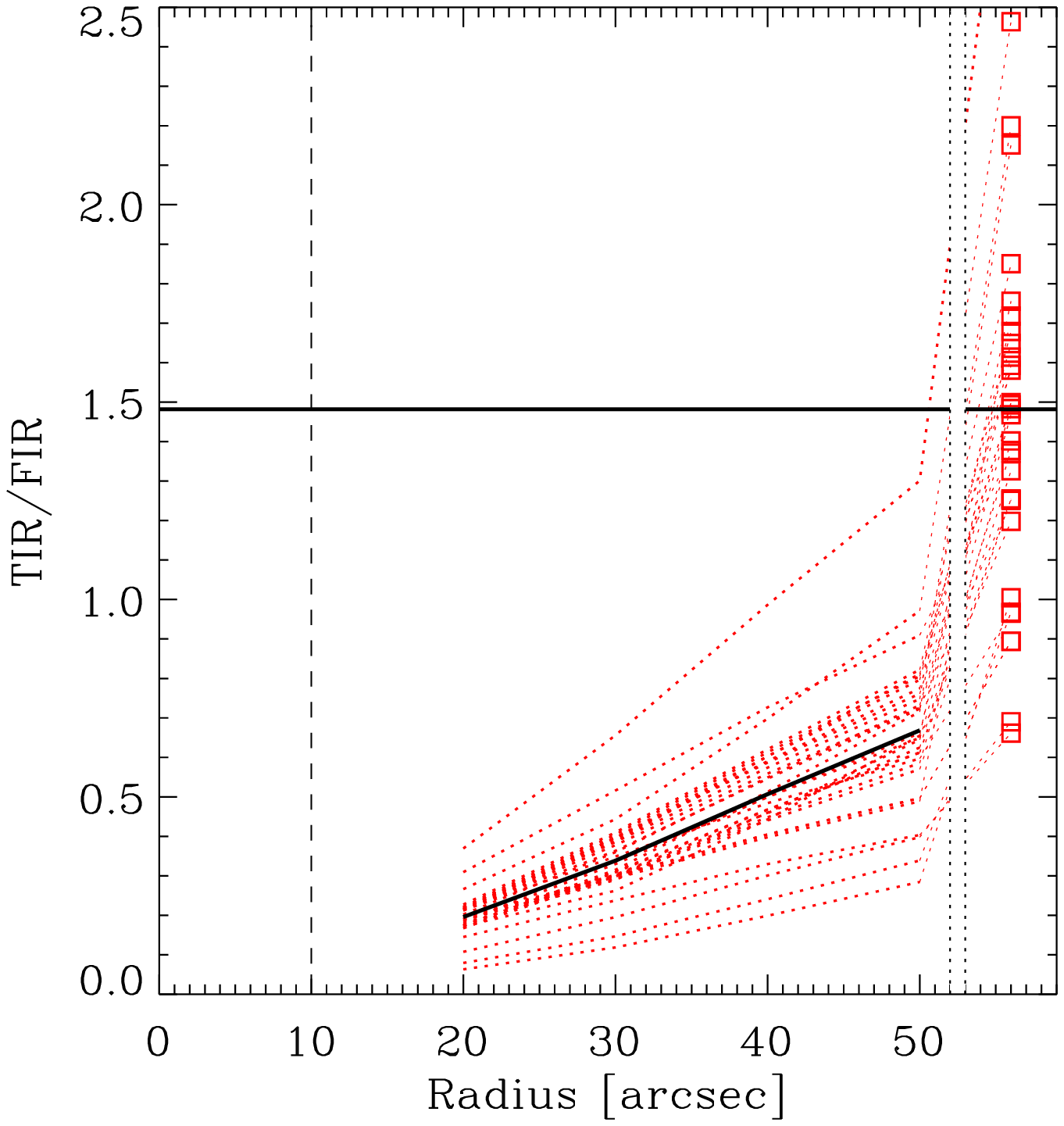}
\caption{
The ratio between TIR and FIR as a function of the aperture radius.
{ Since FIR values are measured from {\iras}, we fix them and
measured the TIR/FIR ($F_{\rm TIR}/F_{\rm FIR}$) ratio as a function
of aperture radii on {\akari} images.
}
Dotted lines represent the TIR/FIR ratio
as a function of aperture radius, and the symbols represent 
convergence values of TIR/FIR for each sample.
The thick solid broken line shows the average value.
The horizontal thick line shows the convergence value of TIR/FIR at large
aperture limit.
Correspondence between the data values at $50''$ radius and convergence
values is indicated also by the dotted lines, with a vertical gap to clarify
that there is a large radius range omitted between them.
{ The leak of flux to the sidelobes of the {\akari} PSF is again not corrected.
}
}\label{fig:tir_fir_radial}
\end{figure*}

\section{Discussion}\label{sec:discussion}

\subsection{IRX as a function of aperture radius}
\label{subsec:irx_radii}

In the above discussion, we have rather focused on the aperture effect of
the UV photometry on the IRX-$\beta$ relation.
Since we have the \akari\ diffuse image map, it would be also interesting to
explore the radial dependence of the IRX values.
Here we estimate the IRX-$\beta$ relation as a function of aperture radius
both on the \galex\ and \akari\ images.
For this purpose, we have put the same elliptical aperture homologous to 
the original \iue\ one, both on these images for each galaxy.

{
Since the angular resolution of the \akari\ diffuse image map is $\sim 30\mbox{--}50''$ at 
{\it WIDE-S} band, smaller aperture radii than $20''$ are not meaningful.
We started an aperture radius from $20''$ which is similar to the \iue\ aperture
radius, and then estimated the IRX with $30''$, $50''$, and $70''$ radii.
We should also take into account the different of the PSFs at FIR and FUV.
It is pointed out that the synthesized PSFs of the {\akari} diffuse map are more
extended than the previous estimates, with a significant flux leak to the sidelobes
(Arimatsu et al., in preparation).
Since it is still difficult to evaluate this effect precisely at this moment, we instead convolved the
{\galex} FUV images with the difference of the angular resolutions of {\galex} and {\akari}, 
we have convolved the {\galex} FUV images with the synthesized PSF of {\akari} FIS 140~$\mu$m 
band and measured the IRX within the same aperture at FIR and FUV bands.
}
The result is shown in Figs.~\ref{fig:galex_akari_2band} and \ref{fig:five_aperture_2band}.

It is very clearly seen that the IRX values of central regions of the sample are 
much lower than the relation obtained from the whole galaxies.
This is mainly because of the difference of the surface brightness profiles at
FUV and FIR wavelengths.
In general, the FUV light of galaxies is much more concentrated than that of FIR.
Then, the IRX values are well below the convergence values within $15''$ aperture,
and then moves up the diagram with increasing aperture radius, as clearly seen in
Fig.~\ref{fig:five_aperture_2band}.
Even with the $50''$ aperture radius, the IRX values are still far from the convergence
values for some galaxies, 
{ even if we take into account the difference of the PSFs.}
This implies that the starburst regions of the M99's original sample galaxies emit
less energy at FIR compared to the whole galaxy scale.

{}To see this more clearly, we show the fraction of TIR flux as a function of aperture
radius in Fig.~\ref{fig:tir_frac}.
Though the values are largely dispersed, on average the fraction contained in $20''$
aperture radius\footnote{This is practically the smallest meaningful aperture for \akari\ diffuse map,
as mentioned in the main text. 
Because of this limited angular resolution, we did not try to discuss this issue based on
the physical radius of the sample galaxies.} is $8.8\pm 3.1$~\%.
Namely, { though we should keep in mind that the correction factor of the
flux leak could be large at small aperture radii, at least we can claim that
less than $\mbox{a few} \times 10$~\% of the TIR light is associated to the central starburst regions
in the sample, but comes from outer part of galaxies. 
}

It will be interesting to simulate the IRX as a function of redshift, taking into account 
a realistic beam size and detection limit for current/future observational facilities to
interpret the high-$z$ IRX-$\beta$ relation.

\subsection{Bolometric correction for the total dust emission}
\label{subsec:dust_BC}

Since the original IRX-$\beta$ relation was based on \iras\ observation, we should 
convert the formula to that of TIR and FUV. 
The assumption of M99 was, as stated above, the ratio of TIR/FIR [the bolometric
correction of dust emission, $\rm BC(FIR)_{Dust}$: see eq.~(\ref{eq:irx_extinction_general})] 
is $\sim 1.4$.
Many authors used this value to rescale the relation with this factor to obtain
the IRX defined by $\log \left( F_{\rm TIR}/F_{\rm FUV} \right)$ \citep[e.g.][]{kong04}.
Then, it is important to examine the value of $\rm BC(FIR)_Dust$ by new IR data.

By making use of the \akari\ FIS diffuse map, we examine the ratio of
TIR and \iras-based FIR.
We calculated a ratio of TIR/FIR ($F_{\rm TIR}/F_{\rm FIR}$), shown in Fig.~\ref{fig:tir_fir_radial}.
In Fig.~\ref{fig:tir_fir_radial}, the TIR/FIRs among the sample galaxies and 
its convergence values are presented.
In Fig.~\ref{fig:tir_fir_radial}, dotted lines represent the TIR/FIR ratio,
and the symbols represent convergence values of TIR/FIR for each sample.
The thick solid broken line shows the average value.
The horizontal thick line shows the convergence value of TIR/FIR at large
aperture limit.
{}From this analysis, we found that the TIR/FIR ratio converges to $1.44 \pm 0.574$, 
which is very close to the ratio originally assumed by M99.
By using a formula of \citet{dale02},  \citet{kong04} discussed that 
TIR/FIR is $\sim 1.5$, which is also quite close to our value.

\subsection{Comparison of IRX-$\beta$ relations between various samples}

\begin{figure*}[tb]
\centering\includegraphics[width=0.7\textwidth ]{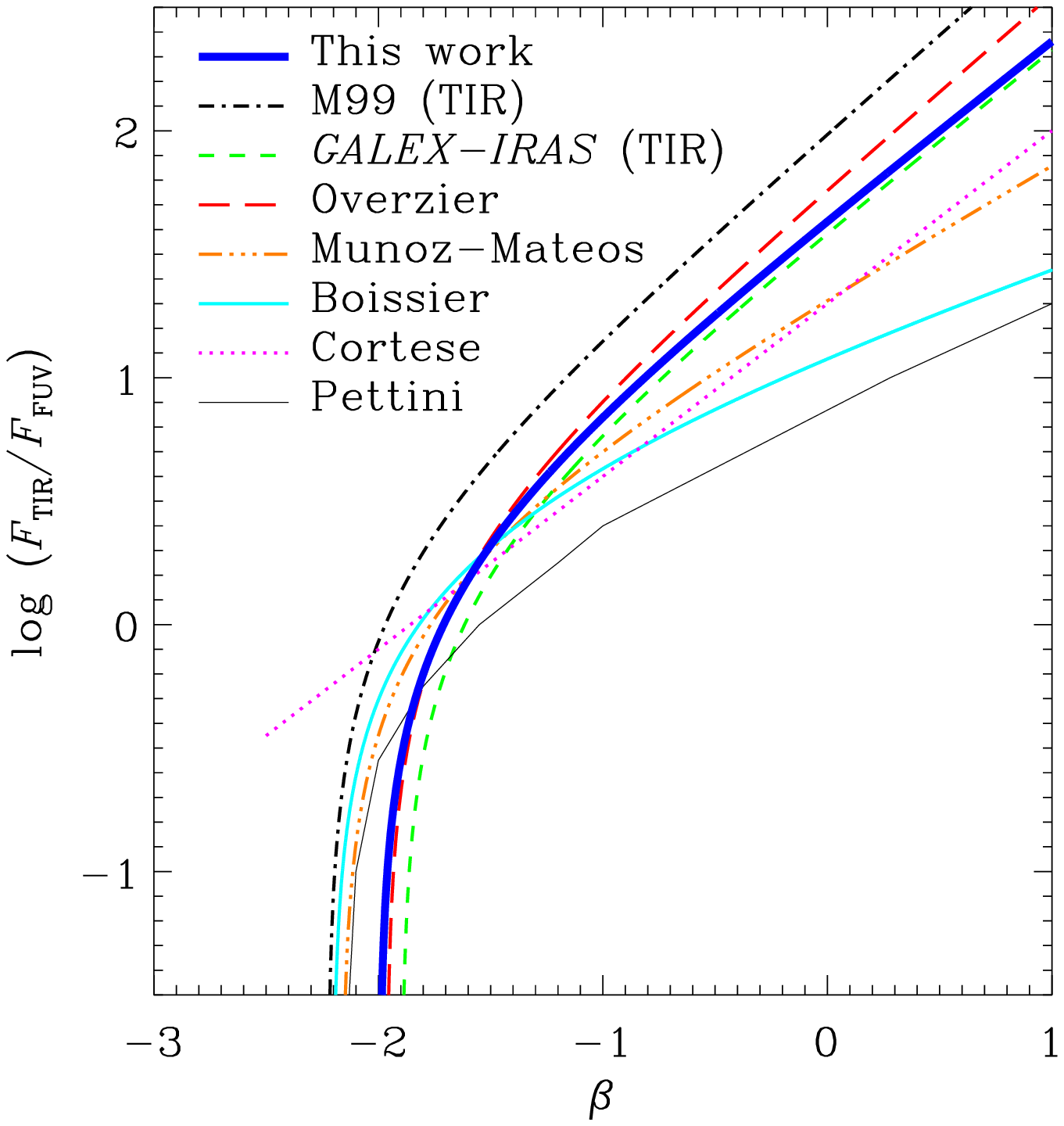}
\caption{
Comparison of various IRX-$\beta$ relations in the literature.
Thick solid curve: this work; 
dot-dashed curve: M99's curve converted from $F_{\rm FIR}/F_{1600}$ to
$F_{\rm TIR}/F_{1600}$;
dashed curve: curve for the M99's sample but with new measurements at UV
and converted for IR flux, i.e., $F_{\rm TIR}/F_{\rm FUV}$.
long dashed curve: same as $F_{\rm TIR}/F_{\rm FUV}$ but with measurements 
by \citet{overzier11};
three-dot-dashed curve: the IRX-$\beta$ relation for optically selected sample \citep{munoz_mateos09};
solid curve: similar to \citet{munoz_mateos09} but with older data \citep{boissier07};
dotted curve: a (bisector) linear fit by \citet{cortese06};
and thin broken solid curve: the IRX-$\beta$ relation from the SMC extinction curve \citep{pettini98}.
}\label{fig:irx_beta_comp}
\end{figure*}

We now compare our new formula with previous studies on the IRX-$\beta$ relation 
with various sample selection.
We summarize the comparison in Fig.~\ref{fig:irx_beta_comp}.
\citet{cortese06} compiled a sample of optically selected galaxies observed by {\galex}
and analyzed the IRX-$\beta$ relation, as well as the {\iue}-selected UV starbursts.
They have converted the FIR flux to the TIR one by the formula of \citet{dale01}.
Since the formula of \citet{dale01} is defined for a wavelength range of $\lambda = 3\mbox{--}
1000\;\mu$m,  it gives a slightly larger value than our formula, but the difference is
small enough compared with the intrinsic scatter and measurement errors \citep{takeuchi05b}.
{}To describe the average trend, they fit a simple linear function instead of the form of 
eq.~(\ref{eq:irx_beta_meurer_general}).
In their analysis, \citet{cortese06} found that the optically selected sample distribute 
systematically lower than starbursts on the IRX-$\beta$ plane.
They suggested qualitatively that this is partially due to the \iue\ aperture, as well as
some physical reason related to the star formation activity.
Our formula passes through the peak of the distribution of their optical sample, and 
agrees very well with their linear fit at $\beta \ga -1$.

\citet{boissier07} selected 43 nearby late-type spiral galaxies optically and performed 
a detailed analysis on the dust attenuation.
The sample was cross-matched with \galex\ Atlas of Nearby Galaxies \citep{gil_de_paz07} 
for UV and \iras\ HIRES images\footnote{URL: http://irsa.ipac.caltech.edu /IRASdocs/hires$\_$over.html.} 
for IR.
The same as \citet{cortese06}, \citet{boissier07} also used the formula of \citet{dale01} to
compute the TIR flux from \iras\ flux.
For the fitting of the IRX-$\beta$ relation, they used a form 
\begin{eqnarray}
 \log \left( \frac{F_{\rm TIR}}{F_{\rm FUV}} \right) = \log \left[ 10^{a_0 + a_1 ({\rm FUV} - {\rm NUV})} - a_2 \right] \;, 
\end{eqnarray}
which has one more free parameter than eq.~(\ref{eq:irx_beta_akari}).
Their sample also distributed below the M99 relation (their Fig.~1).
Their IRX-$\beta$ relation has a higher IRX at smaller $\beta$ and lower IRX at larger $\beta$,
but since there are very few or no data at $\beta < -2$, the discrepancy on this side is not
crucial.
At the larger $\beta$ side, actually there is a large dispersion in optically selected spiral
galaxy sample in \citet{boissier07}. 
This may be the reason of their flat slope.
Indeed,  \citet{munoz_mateos09} used the SINGS galaxy sample \citep{kennicutt03} for 
a similar analysis and obtained an IRX-$\beta$ relation with steeper slope at large $\beta$ 
side.
The new relation of \citet{munoz_mateos09} is in a very good agreement with our result
at $\beta > -1.5$.
Thus, we can safely say that our new IRX-$\beta$ formula determined by the original UV 
starburst sample of M99, also describes the attenuation of optical sample very well.
However, we should keep in mind that there is a considerable scatter around the relation
obtained by fitting, as these authors also cautioned.

Next, we also examine if eq.~(\ref{eq:irx_beta_akari}) would apply to the IR-selected samples.
As we have already seen, the behavior of IR-selected galaxies on the IRX-$\beta$ plane is 
quite complicated.
These galaxies have two branches of distribution, one locates below the M99 relation and
the other spreads well above the line \citep[e.g.,][]{buat05,buat10,takeuchi10}.

\citet{goldader02} first discovered that the population distributing above the M99 line is
LIRGs or ULIRGs. 
\citet{takeuchi10} examined this aspect intensively and found that the population distributing 
lower than the M99 line is quiescent star-forming galaxies, and confirmed that the one above 
the M99 line is LIRGs or ULIRGs, by evaluating the 60-$\mu$m luminosity of \akari\ FISBSC 
sample.

\citet{buat05} presented an impressive IRX-$\beta$ plots for both \galex\ NUV- and 
\iras\ 60~$\mu$m-selected samples.
The quiescent branch of IR-selected galaxies distribute in almost the same way as
the UV-selected sample on this plane, i.e., below the M99's IRX-$\beta$ relation 
but in parallel with it.
More active branch galaxies are spread over the M99 relation.
We see that the former is well described by eq.~(\ref{eq:irx_beta_akari}) without significant
bias.
However, since the distribution of (U)LIRGs are very different from that of quiescent 
galaxies, our new relation cannot apply to these active galaxies.

As for high-$z$ galaxies, still it is not very easy to obtain both UV and IR data, but gradually
direct measurements are reported recently.
\citet{siana09} showed the IRX-$\beta$ measurements for two such objects, MS1512-cB58 
($z = 2.724$) and the Cosmic Eye ($z = 3.074$), both of which are Lyman-break galaxies 
(LBGs), i.e., UV-luminous starbursts.
They found that they locate significantly below the original M99 relation.
\citet{reddy10} also discussed the IRX-$\beta$ relation for both UV- and IR-luminous
galaxies at $z \sim 2$.
They found that their ``typical'' $z \sim 2$ galaxies do follow the M99 relation (but with
a large scatter), and ULIRGs at the same redshift locate well above it similarly to their Local 
counterparts.
Also they argue that young galaxies at these redshifts distribute below the M99 line.
The young galaxies seem to follow the IRX-$\beta$ relation expected for the SMC
extinction curve \citep{pettini98}, also suggested by \citet{siana09}.
Comparing their results with our new IRX-$\beta$ relation (see Fig.~\ref{fig:irx_beta_comp}),
we would say the $z \sim 2$ typical galaxies of \citet{reddy10} locate above the new 
IRX-$\beta$ relation, while the young galaxies follow it.
As for lensed $z \sim 3$ galaxies in \citet{siana09}, cB58 is approximately on the new relation,
while the Cosmic Eye may be still below the relation.
It would be very interesting to make a statistical analysis when a large sample of galaxies
at these redshifts will be available, to discuss whether evolutionary effect really would 
take place.
{ We should note that the IRX-$\beta$ relation of the central starburst regions of
our sample locates well below the M99's original relation and our new result, but 
more similar to the Pettini's relation (see top-left panel of Fig.~\ref{fig:five_aperture_2band}).
Since M99 originally selected the sample to have a central starbursts, we might speculate, for example, 
that the properties of high-$z$ star-forming galaxies 
could be similar to that of a pure starburst region.
Further investigation in this direction will be interesting.
}

\citet{kong04} suggested a systematic sequence of the IRX-$\beta$ relation from 
lower to higher along with the birthrate parameter $b$
\begin{eqnarray}
  b \equiv \frac{\mbox{SFR}(t_0)}{\langle \mbox{SFR}\rangle}
\end{eqnarray}
where $t_0$ is a supposed age of a galaxy \citep{kennicutt94}.
They claimed that $b$ is related to the attenuation as
\begin{eqnarray}
  A_{\rm FUV} = 3.87 + 1.87 (\beta_{\rm GLX} + 0.40 \log b ) \;.
\end{eqnarray}
In \citet{kong04}, the original M99 relation corresponds to the extremely bursting
group, i.e., $b > 3.0$, not very consistent with the properties of M99's UV-luminous 
starburst sample.
We see that our new IRX-$\beta$ relation corresponds to the intermediate $b$ galaxies 
similar to normally star forming galaxies, now consistent with the original M99 sample.

\citet{overzier11} independently made essentially the same analysis with this study
but with their own \galex\ photometry and {\iras}-based IR flux estimation
with a conversion formula of \citet{dale01}, with {\it Spitzer} measurement for
a subsample of 12 galaxies
{
Their formula is, as we expect, not very different from our result [eq.~(\ref{eq:irx_beta_akari})], 
but slightly deviated upward than ours at large $\beta$ and high IRX regime.
They have measured the UV flux up to $2.5 \mbox{--} 5$ times
Kron radius of galaxies, their converged values are on average twice larger than
the original values of M99.
As we have shown, the \iue\ flux is underestimated with a factor of 2.5 on average, 
and sometimes the underestimation is of an order of magnitude.
Hence, though they discuss that the quiescent IR galaxies in \citet{buat10} locate lower than their
line, for example, this might be due to the UV flux still missed.
}


\section{Conclusion}\label{sec:conclusion}

The relation between the ratio of infrared (IR) and ultraviolet (UV) flux densities
(the infrared excess: IRX) and the slope of the UV spectrum ($\beta$) of 
galaxies plays a fundamental role in the evaluation of the dust attenuation of star forming 
galaxies especially at high redshifts. 
{M99 first introduced a very useful method to estimate the dust attenuation by using
the relation between the FIR-to-FUV flux ratio and the UV slope, the IRX-$\beta$ relation. 
However, subsequent studies revealed a dispersion
and/or systematic deviation from the original IRX-$\beta$ relation of M99, 
depending on sample selection.
}
We reexamined the original IRX-$\beta$ relation proposed by M99 by measuring the far- and 
near-UV flux densities of the original sample galaxies used by M99\nocite{meurer99} with 
{\galex} imaging data.
We summarize our conclusion as follows: 
\begin{enumerate}
\item The UV flux densities of the galaxies used in M99\nocite{meurer99} were 
significantly underestimated because of the small aperture of {\iue} which was 
$10'' \times 20''$. 
Consequently, the IRX values were lower than that in M99\nocite{meurer99},
which lead the IRX-$\beta$ relation shifted downward on the plot.
\item The aperture effect affected not only the IRX but also $\beta$, because 
the curve of growth of the flux densities is a function of aperture
which depends strongly on the wavelength.
\item Correcting the aperture effect, we obtained a new IRX-$\beta$ relation for UV-selected starburst 
galaxies for the same sample with M99\nocite{meurer99}.
\item The original relation was obtained based on \iras\ data. 
Since \iras\ bands cover only a wavelength range of $\lambda = 42\mbox{--}122\;\mu$m, 
the FIR flux calculated from \iras\ flux does not represent the total IR emission from dust.
We used data from \akari\, which has a much wider wavelength coverage especially toward
longer  wavelengths, and obtained an appropriate IRX-$\beta$ relation for the data with 
total dust emission. 
\item We also estimated the IRX-$\beta$ relation of the sample as a function both on
the \galex\ and \akari\ images.
We found that the IRX-$\beta$ relation of the central starburst regions of the sample 
locates significantly below the relation obtained for the whole galaxy regions.
This means that the extended dust emission is important to evaluate various global
properties of galaxies, 
{ though we need to know the behavior of the PSFs of AKARI diffuse map more precisely to
evaluate more quantitatively.
} 
\item In many previous studies, authors use a conventional conversion factor (bolometric correction: BC) 
of $\sim 1.4$ to obtain the total IR flux from dust.
We examined the BC of dust emission and found that indeed the value was $1.44 \pm 0.574$
on average.
\item Our new relation is consistent with most of the preceding results on the IRX-$\beta$ relation
for samples selected at optical and UV, though there is a significant scatter around it.
We also found that even the quiescent class of IR galaxies follow this new relation, though
luminous and ultraluminous IR galaxies distribute completely differently. 
\item The IRX-$\beta$ relation for the central starburst regions of M99's sample galaxies
is very similar to the relation proposed by \citet{pettini98}, based on the SMC attenuation
curve.
This may imply interesting property of high-$z$ star-forming galaxies. 
\end{enumerate}
In this work, we constructed a proper formula for a category of galaxies which are luminous
in UV and actively forming stars.
We found that this formula also applies to various class of moderately star-forming galaxies,
like optically selected or non-luminous IR selected ones.
This will be a more firm basis than before when one needs to correct dust extinction of high-$z$ galaxies. 
However, we should note the large intrinsic scatter on the IRX-$\beta$ plane, especially
the (U)LIRG population which does not obey the IRX-$\beta$ relation at all.
Many issues and problems still remain to be solved for understanding this relation.

\bigskip

First of all, we thank the anonymous referee for her/his extremely careful and 
constructive comments and suggestions that improved this article significantly.
This work is based on observations with {\akari}, a
JAXA project with the participation of ESA.
We thank Yasuo Doi for preparing the \akari\ FIS diffuse map for our 
sample galaxies, and Ko Arimatsu for providing the new synthesized PSF
of the map.
We also thank V\'{e}ronique Buat, Denis Burgarella, 
Agnieszka Pollo, Ryosuke Asano, Fumiko Nagaya, Mai Fujiwara, Daisuke Yamasawa, 
and Takashi Kozasa for fruitful discussions and comments. 
We are grateful to Takako T.\ Ishii for developing the basis of
the photometric software with IDL. 
TTT has been supported by
Program for Improvement of Research Environment for Young
Researchers from Special Coordination Funds for Promoting Science
and Technology.
TTT and AKI have been also supported by the Grant-in-Aid for the Scientific Research
Fund (20740105, 23340046, 24111707: TTT, 19740108: AKI) commissioned by the Ministry of Education, Culture,
Sports, Science and Technology (MEXT) of Japan.  
TTT, FTY, AI, and KLM have been partially
supported from the Grand-in-Aid for the Global COE Program ``Quest
for Fundamental Principles in the Universe: from Particles to the
Solar System and the Cosmos'' from the MEXT.

Facilities: 
\facility{{\galex}}, 
\facility{{\iue}},
\facility{{\iras}},
\facility{{\akari}}

\appendix

\setcounter{figure}{0}

\section{A. Formulation of \citet{meurer99}}\label{sec:meurer}

Here, we review the original formulation of the IRX-$\beta$ relation
by M99\nocite{meurer99}.
\begin{equation} \label{eq:irx_def}
  {\rm IRX_{1600}} \equiv   \frac{F_{\rm FIR}}{F_{\rm 1600}}
  =\frac{\displaystyle F_{\rm Ly\alpha}+\int^{\infty}_{\mbox{\tiny 912\AA}} f_{\rm \lambda ',0}
  (1-10^{-0.4A_{\rm \lambda '}})d\lambda '}{\displaystyle F_{1600,0}10^{-0.4A_{1600}}}
\displaystyle\left (\frac{F_{\rm FIR}}{F_{\rm bol}}\right)_{\rm Dust} \; .
\end{equation}
where $f_{\rm \lambda ', 0}$ is the unattenuated flux density of the emitted
spectrum, $A_{\rm \lambda '}$ is the net absorption in magnitudes by dust as 
a function of wavelength, and $F_{\rm Ly\alpha}$ is the Ly$\alpha$ flux.
In eq.~(\ref{eq:irx_def}), it is assumed that the ionizing photons does not affect 
dust heating significantly, namely they are not absorbed directly by dust 
grains.
\footnote{M99\nocite{meurer99} 
argue that under the starburst population model, typically only $\la 10$~\% of the
bolometric luminosity of starbursts is emitted below 912~\AA\ \citep{leitherer95}, 
and the fraction of ionizing photons that is directly absorbed by dust grains is 
estimated to be $\sim 25$~\% \citep{smith78}.
However, the direct absorption of Lyman photons by dust grains in ionizing regions
has been first discussed by \citet{petrosian72} and extensively by \citet{inoue02}, 
who found that the fraction of ionizing photons directly absorbed by dust is 
significantly larger than these values.
{However, even if all the ionizing photons were absorbed by dust, they would 
still not drastically affect the conclusions.
Hence, the formulation of M99 will remain valid.
}
}. 
The Ly$\alpha$ flux $F_{\rm Ly\alpha}$ was produced by cascade of photons with wavelengths
shorter than 912~\AA, which finally reach the transition $n: \; 2 \rightarrow 1$.
If the ionized gas in Case B, Ly$\alpha$ is resonantly scattered repeatedly by hydrogen atoms
and finally absorbed by dust.
Thus, the numerator of eq.~(\ref{eq:irx_def}) corresponds to the total flux absorbed by dust,
consisting of two terms, the first is the heating by Ly$\alpha$ and the second is by nonionizing photons.
The denominator in eq.~(\ref{eq:irx_def}) represents the observed 1600~\AA\ flux density 
which is attenuated by dust.
Namely, 
\begin{eqnarray}
  F_{\rm bol} &=& F_{\rm Ly\alpha} + \int^{\infty}_{912\;{\rm \AA}} f_{\rm \lambda ',0}
  (1-10^{-0.4A_{\rm \lambda '}})d\lambda '  \; , \\
  F_{1600}&=&F_{1600,0}10^{-0.4A_{1600}}\;,  \\
  \displaystyle \left( \frac{F_{\rm FIR}}{F_{\rm bol}}\right)_{\rm Dust} &=&
  \frac{\mbox{Flux of dust emission within the wavelength range from $42$ to $122\;\mu$m}}{\mbox{Total flux of  dust emission}}.
\end{eqnarray}
 
Most of the photons absorbed by dust are contribution from massive stars.
Then, by assuming that most of the radiation from massive stars is UV, 
we can substitute $A_{\lambda}$ with $A_{1600}$ in eq.~(\ref{eq:irx_def}) as
\begin{eqnarray}\label{eq:irx_ap1600}
  {\rm IRX_{1600}}
  = (10^{0.4A_{\rm 1600}}-1)\frac{F_{\rm Ly\alpha}+\int^{\infty}_{912\;{\rm \AA}} f_{\rm \lambda ',0}
  d \lambda '}{F_{1600,0}} 
  \displaystyle\left (\frac{F_{\rm FIR}}{F_{\rm bol}}\right)_{\rm Dust} \;.
\end{eqnarray}

Further, we rearrange eq.~(\ref{eq:irx_ap1600}) as follows. 
\begin{eqnarray}\label{eq:bc}
 {\rm IRX_{1600}} &=& (10^{0.4A_{1600}}-1) B \; , \label{eq:irx_bc} \\
 B &=& \displaystyle {\rm \frac{BC(1600)_*}{BC(FIR)_{Dust}}} \;
\end{eqnarray}
The subscript asterisk means it is related to stellar emission, and subscript
Dust means it is for dust emission.
Here, 
\begin{eqnarray}
 &&{\rm BC(1600)_*} = \displaystyle \left (\frac{F_{\rm Ly\alpha}+\int^{\infty}_{\rm 912\AA}
  f_{\rm \lambda ',0}d\lambda '}{F_{1600,0}}\right )\\
 &&{\rm BC(FIR)_{Dust}} = \displaystyle\left (\frac{F_{\rm bol}}{ F_{\rm FIR}}\right)_{\rm Dust} \;.
\end{eqnarray}
We assume 
\begin{eqnarray}\label{eq:bc_ratio}
  B = \frac{\rm {BC(1600)_*}}{\rm BC(FIR)_{Dust}} = \mbox{const.} 
\end{eqnarray}
On one hand, $\rm{BC(1600)_*}$ can be approximated as a constant if we fix the initial mass 
function (IMF) and a constant star formation rate.
If we adopt the Salpeter IMF \citep{salpeter55} with an upper mass limit of $100\;M_\odot$, 
we have $\rm{BC}(1600)_*=1.66\pm 0.15$ (\nocite{meurer99}M99).
The uncertainty is calculated based on the variation of $\rm {BC}(1600)_*$ depending on
the burst duration. 
On the other hand, $\mbox{BC(FIR)}_{\rm Dust}$ is the ratio between the total IR flux and 
the FIR flux. 
Empirically from the \iras\ observation \citep{rigopoulou96}, M99 adopted a value of 
BC(FIR)$_{\rm Dust}=1.4\pm 0.2$, which leads $B = 1.19\pm0.20$. 
Putting them together with eq.~(\ref{eq:irx_bc}), we obtain
\begin{eqnarray}
  {\rm log(IRX_{1600})} &=& {\rm log} (10^{0.4A_{1600}}-1) + \log B \nonumber \\
  &=&{\rm log}(10^{0.4A_{1600}}-1) + 0.076 \pm 0.044 \;.
\end{eqnarray}
which is eq.~(\ref{eq:irx_extinction}) in the main text, 
the basic relation between $\mbox{IRX}_{1600}$ and $A_{1600}$. 

\section{B. Examination of the aperture effect in \akari\ FIS Bright Source Catalog}
\label{sec:akari_diffuse}

\begin{figure*}[t]
 \centering\includegraphics[width=0.7\textwidth]{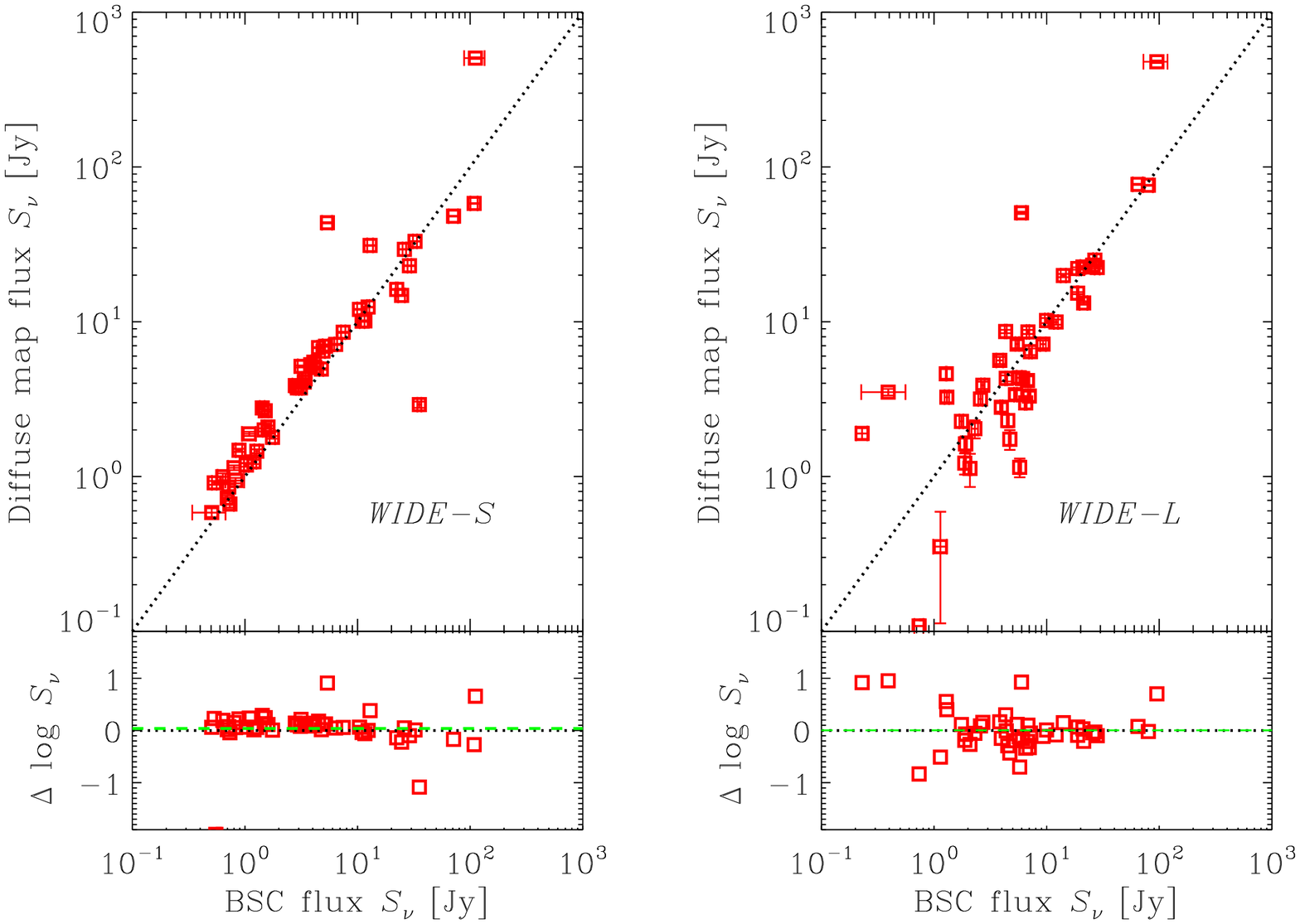}
\caption{
Comparison between the \akari\ FISBSC flux density and that measured from 
the diffuse map at {\it WIDE-S} (left) and {\it WIDE-L} (right).
Diagonal lines represent $y = x$.
The lower panels show the logarithmic deviation of data from the diagonal lines.
Horizontal dashed lines are the average deviation of data at each band.
}\label{fig:akari_diffuse}
\end{figure*}

The \akari\ FISBSC was produced by a dedicated pipeline for point source 
extraction \citep{yamamura10}.
However, for very extended nearby galaxies, the point source assumption is
not valid anymore.
{}To examine this ``aperture'' effect in \akari\ FIS flux density, we measured
IR flux density from \akari\ FIS diffuse map.
We performed this procedure by using both {\sc SExtractor} \citep{bertin96} and our own software, 
and both yielded more or less the same result. 
Since \citet{yuan11} used {\sc SExtractor}, we also present flux density 
given by {\sc SExtractor} for the diffuse map. 
We show the comparison between flux densities from FISBSC and diffuse map
for M99 sample in Fig.~\ref{fig:akari_diffuse}.

Basically these flux densities agree well, because M99 put a constraint to have
galaxies with a small angular diameter in their sample selection.
However, even so, the BSC flux density turned out to be underestimated for
non-negligible number of galaxies.
The horizontal dashed lines represent the average difference between BSC 
and diffuse-map flux densities.
For both {\it WIDE-S} and {\it WIDE-L}, the average is positive.
\citet{yuan11} performed a careful comparison of BSC flux density with other
well-understood database and concluded that this is due to the underestimation
of the BSC and not the overestimation in diffuse map photometry.
Thus, we use the diffuse-map based IR flux densities to have the final IRX-$\beta$
relation fitting.

\end{document}